\documentclass{jfm}
\pdfoutput=1 
\usepackage{graphicx}
\usepackage{newtxtext}
\usepackage{newtxmath}
\usepackage{natbib}
\usepackage{amsmath}
\usepackage{array}
\usepackage{upgreek}
\usepackage{comment}
\usepackage{textcomp}
\usepackage{pifont}
\usepackage{hyperref}
\usepackage{booktabs}
\usepackage[dvipsnames]{xcolor}

\usepackage{soul}

\hypersetup{
    colorlinks = black,
    urlcolor   = blue,
    citecolor  = black,
}

\newcommand{\RomanNumeralCaps}[1]
\linenumbers
\newcommand{\comm}[1]{}

\shorttitle{Interaction of a rising bubble with a vertical wall}
\shortauthor{P. Shi, J. Zhang, J. Magnaudet}

\title{Lateral migration and bouncing of a deformable bubble rising near a vertical wall. Part 1. Moderately inertial regimes}

\author{Pengyu Shi\aff{1,3}\corresp{\email{pengyu.shi@toulouse-inp.fr}}, 
Jie Zhang\aff{2} \corresp{\email{j\_zhang@xjtu.edu.cn}}
\and Jacques Magnaudet\aff{1}
	\corresp{\email{jmagnaud@imft.fr}}}
	
\affiliation{\aff{1}Institut de M\'ecanique des Fluides de Toulouse (IMFT), Universit\'e de Toulouse, CNRS, Toulouse, France
\aff{2}State Key Laboratory for Strength and Vibration of Mechanical Structures, School of Aerospace, Xi'an Jiaotong University, Xi'an, PR China
\aff{3}Helmholtz-Zentrum Dresden Rossendorf, Institute of Fluid Dynamics, 01328 Dresden, Germany}

\begin{document}
\maketitle
\begin{abstract}
The buoyancy-driven motion of a deformable bubble rising near a vertical hydrophilic wall is studied numerically. We focus on moderately inertial regimes in which the bubble undergoes low-to-moderate deformations and would rise in a straight line in the absence of the wall. Three different types of near-wall motion are observed, depending on the buoyancy-to-viscous and buoyancy-to-capillary force ratios defining the Galilei ($Ga$) and Bond ($Bo$) numbers of the system, respectively. For low enough $Ga$ or large enough $Bo$, bubbles consistently migrate away from the wall. Conversely, for large enough $Ga$ and low enough $Bo$, they perform periodic near-wall bounces. At intermediate $Ga$ and $Bo$, they are first attracted to the wall down to a certain critical distance, and then perform bounces with a decreasing amplitude before stabilizing at this critical separation. Periodic bounces are accompanied by the shedding of a pair of streamwise vortices in the wake, the formation of which is governed by the near-wall shear resulting from the no-slip condition. These vortices provide a repulsive force that overcomes the viscous resistance of the fluid to the departing motion, making the bubble capable of returning to the region where it is attracted again to the wall. Although periodic, the shedding/regeneration cycle of these vortices is highly asymmetric with respect to the lateral bubble displacements, vortices being shed when the gap left between the bubble and the wall reaches its maximum, and reborn only when this gap comes back to its minimum.  
\end{abstract}
	
	

\section{Introduction}
\label{sec:intro}
Wall-bounded bubbly flows are widespread in engineering processes. In nuclear reactors, boiling water creates bubbles on heated surfaces, which affects the reactor efficiency and poses safety challenges. In hydrogen production through water electrolysis, bubbles emerge in the electrolyte due to the formation of hydrogen at the cathode and oxygen at the anode. Efficient removal of these bubbles from the electrodes surface is vital for the performance of the system. In froth flotation processes, particularly in reflux flotation cells, an inclined channel section is deliberately introduced to enhance the separation of bubbles from the slurry, thus mitigating the loss of attached hydrophobic particles. In these processes and in many others, accurately predicting the distribution of bubbles near walls is essential. However, the complexity of the wall-bubble interaction processes challenges this prediction \citep{2008_Yin, 2011_Takagi, 2013_Lu}.
	
At the local scale, the primary step in the understanding of these interactions consists in considering the rise of an isolated bubble in a quiescent fluid partially bounded by a flat wall. However, the orientation of this wall dictates to a large extent the main physical ingredients governing the interaction sequence, and therefore the fate of the bubble. In the presence of a horizontal wall, the problem exhibits an axial symmetry, provided that the bubble is small enough to rise in a straight line. Interactions usually manifest themselves in a series of damped near-wall bounces. This bouncing sequence is mostly governed by the time-dependent variations of the bubble shape with the distance to the wall, which yield variations in both the bubble surface energy and the kinetic energy of the surrounding fluid, and by lubrication effects in the gap \citep{1997_Tsao,2009_Zenit,2013_Zawala, 2014_Klaseboer, 2014_Kosior}. In contrast, the configuration where the bubble rises close to a vertical wall is intrinsically three-dimensional. Bubble-wall interactions are then driven by non-axisymmetric effects, be they inertial or viscous in nature. Moreover, the presence of a wake behind the bubble plays a key role in the interaction process whatever the hydrodynamic regime, leading to different styles of path according to the relative magnitude of inertial, viscous and capillary effects \citep{2002_devries, 2003_Takemura, 2007_Zaruba, 2017_Lee, 2020_Zhang, 2022_Yan, 2023_Cai, 2024_Cai, 2024_Estepa}. Intermediate wall inclinations have also been considered, especially with the aim of determining the critical angle beyond which the response of the system transitions from a regime of repeated bounces to another regime in which the bubble slides steadily some distance below the wall \citep{1997_Tsao,2016_Barbosa,  2022_Heydari,2022_Khodadadi}. 
 Additional complexity arises in the presence of surfactants \citep{2020_Ahmed, 2022_Ju}, owing to the Marangoni effect resulting from the localized contamination of the bubble surface. The possible partial wettability of the wall also alters the interaction process \citep{2015_Jeong, 2022_Khodadadi}, as it promotes the formation of a moving three-phase contact line, making the bubble more prone to slide along the wall without detaching from it.\\
\indent In what follows, we concentrate on the configuration where a clean bubble (thus with uniform surface tension) rises close to a vertical, hydrophilic wall. Moreover, we mostly restrict ourselves to moderately inertial regimes in the presence of significant surface tension effects, so that the bubble experiences moderate deformations and would follow a straight vertical path if it were to rise in an unbounded fluid at rest. Highly inertial regimes in which isolated bubbles follow a non-straight path will be examined in a companion
paper. \\
 \indent  To set the scene in more detail, it is useful to come back to the observations and discussion reported by \cite{2003_Takemura} (hereinafter abbreviated to as TM). In their experiments, performed in several silicone oils with nearly-spherical bubbles, these authors identified three distinct interaction regimes, depending on the rise Reynolds number of the bubble, $Re$. For $Re\lesssim35$, with $Re$ based on the bubble equivalent diameter and {\textcolor{black}{actual rise speed in the presence of the wall}}, bubbles were consistently found to migrate away from the wall.
 The mechanism involved lies in the interaction of the wake with the wall, as was first identified for a rigid sphere sedimenting at finite $Re$ close to a vertical wall by \cite{1977_Vasseur}. More specifically, as the sphere translates, a certain amount of fluid is displaced laterally in the wake. At some point, a fraction of this fluid encounters the wall and the latter reacts by generating a small lateral flow directed away from it. Thus, the associated pressure gradient is towards the wall, which, in the presence of finite inertial effects, i.e., at finite $Re$, results in a lateral force directed away from the wall. At low-but-finite Reynolds number, this force decreases as the inverse square of the distance separating the sphere from the wall. \cite{2002_Takemura} extended the prediction of \cite{1977_Vasseur} to spherical bubbles (and drops), showing that the lateral force is proportional to the square of the maximum vorticity at the particle surface. This vortical mechanism is still active at moderate Reynolds number, say $Re=\mathcal{O}(10-100)$, although the lateral force decays more rapidly with the separation distance than predicted in the low-but-finite $Re$ limit.\\
 \indent Beyond $Re\approx35$ and below a second critical Reynolds number close to $65$, uncontaminated nearly-spherical bubbles exhibit a dramatically different near-wall behaviour. Released some distance apart from the wall, they promptly migrate towards it and eventually stabilize a very short distance to it, possibly with some damped oscillations, leaving only a thin interstitial liquid film in the gap. The migration towards the wall merely results from the Bernoulli mechanism that can be inferred from potential flow theory. Indeed, in the inviscid limit, mass conservation implies that, in the bubble's reference frame, the fluid moves faster in the gap than on the opposite side of the bubble. This implies the existence of a pressure minimum in the gap, which results in the migration of the bubble towards the wall. The corresponding lateral force decreases as the fourth power of the inverse of the separation when the latter is of the order of the bubble size or larger \citep{1976_Wijngaarden, 1977_Miloh}. This attractive transverse force is at the origin of the coalescence of weakly-deformed bubbles rising side by side  \citep{1998_Duineveld, 2009_Sanada, 2021_Kusuno}. Although the potential flow model provides only a crude approximation of the actual flow past the bubble when the rise Reynolds number is only a few tens, the measurements of TM indicate that it realistically predicts the transverse force acting on nearly-spherical bubbles rising near a vertical wall as long as the separation is larger than the bubble diameter. For smaller gaps, the above two mechanisms combine in a complex and still unclear manner, resulting in the existence of a wall-normal position very close to the wall at which the total transverse force vanishes. This is the position at which bubbles eventually stabilise.  \\
 \indent Increasing the Reynolds number beyond approximately $65$ while remaining in the range where bubbles only exhibit small deformations (say, up to $Re\approx350$ in water) reveals a third, well distinct interaction scenario (\cite{2002_devries}, TM). In that range, bubbles perform regular bounces very close to the wall while rising, the amplitude of the lateral oscillations being a significant fraction of the bubble size. This oscillatory behaviour may be qualitatively understood by considering that, close to the equilibrium position at which the total transverse force vanishes, this force varies linearly with the wall-normal position. Within this approximation, the transverse force plays a role similar to that produced by a spring, the extremity of which is slightly displaced from its equilibrium position. If damping effects are small, the reaction of the bubble to this force arises primarily through the inertia of the surrounding liquid, in the form of a virtual-mass force, proportional to the amount of liquid displaced by the bubble in the course of its lateral motion. Under such conditions, the bubble-fluid entity behaves essentially as a mass-spring system. Hence, bubbles perform transverse oscillations, the natural frequency of which may be determined once the near-wall variations of the overall transverse force and the virtual mass of the bubble (which depends on its shape and position with respect to the wall) are known. Of course, viscous effects make the fluid resist the lateral bubble displacements, and these effects are responsible for the damping or even the absence of the transverse oscillations when the Reynolds number lies in the intermediate range $35\lesssim Re\lesssim65$. The reasons why no visible damping subsists when $Re\gtrsim65$ in spite of the still existing viscous resistance, is a largely open question, although there is little doubt that the answer lies in the wake dynamics. This dynamics has been explored experimentally with the help of various optical techniques \citep{2002_devries, 2017_Lee, 2023_Cai, 2024_Cai} and through simulations \citep{2020_Zhang, 2022_Yan}. Nevertheless its connection with the shape, rise speed, wall-normal position and transverse velocity of the bubble is still unclear.\\
\indent Building on the above knowledge and open issues, the present investigation aims at providing new insights into the mechanisms governing the various interaction scenarios observed in experiments. More precisely, we aim at making progress on three main questions: How does bubble deformation affect the succession of these scenarios? Which role does the wake dynamics play in the regimes taking place when fluid inertia dominates, especially in the bouncing regime observed for $Re\gtrsim65$ with nearly-spherical bubbles? How do these oscillations of the bubble path affect in turn the behaviour of the wake? \\
\indent The findings analysed in this paper were obtained through a series of high-resolution
simulations covering a significant range of hydrodynamic conditions. Computations were carried out with the open source code \textit{Basilisk} \citep{2015_Popinet} based on the volume of fluid (VOF) approach. The adaptive mesh refinement (AMR) technique implemented in this code made it possible to properly capture the flow in between the bubble and the wall, down to very small gaps. 
The paper is organized as follows. In \S\,\ref{sec:problem_state}, we formulate the problem, specify the range of parameters considered, and outline the numerical approach. Section \ref{sec:path_wake} discusses in detail the characteristics of the path and wake observed in the simulations, which imply the three different hydrodynamic regimes reviewed above. A further analysis of the mechanisms governing the periodic bouncing regime is carried out in \S\, \ref{sec:discuss}. The findings obtained in the course of the investigation are summarized in \S\,\ref{sec:summary}. The paper is supplemented by a series of appendices. Of special significance are appendices \ref{sec:appA} and  \ref{sec:appB} which establish the accuracy of present numerical predictions.
\section{Statement of the problem and outline of the numerical approach}
\label{sec:problem_state}
\subsection{Problem definition}
\begin{figure}
\centerline{\includegraphics[scale=1]{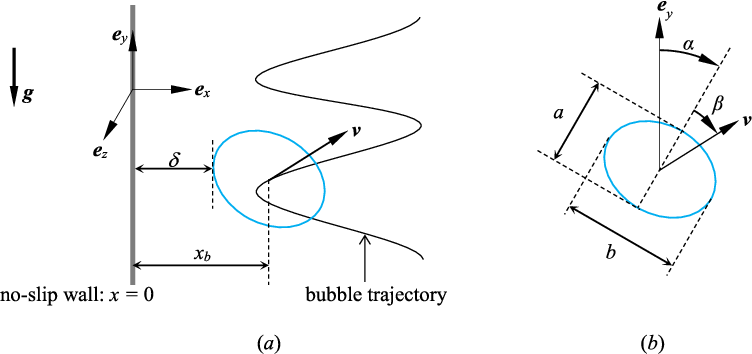}}
\caption{Sketch of the problem and definition of basic quantities.}
\label{fig:problem-state}
\end{figure}

We consider the buoyancy-driven motion of a single gas bubble rising in a stagnant liquid in the presence of a nearby vertical wall. The notations to be used throughout the paper are specified in figure \ref{fig:problem-state}$(a)$. In particular, $\boldsymbol{e}_x$ and $\boldsymbol{e}_y$ are the unit vectors in the wall-normal direction (pointing into the liquid), and along the direction of buoyancy, respectively. Hence, the gravitational acceleration is $\boldsymbol{g}=-g\boldsymbol{e}_y$. The wall is located at $x=0$ and the initial distance from the bubble centre to the wall is $x_0$, such that in Cartesian coordinates ${\bf{x}}=(x,y,z)$ the bubble is released at the position $(x_0, 0, 0)$. The instantaneous gap between the wall and the point of the bubble surface closest to it is $\delta(t)$. During the rise, the position of the geometrical bubble centroid (curved line in figure \ref{fig:problem-state}$(a)$) is ${\bf{x}}_b(t)=(x_b(t), y_b(t), z_b(t))$, while the bubble velocity, which is tangent to the path is $\boldsymbol{v}(t)$.
	
We restrict attention to moderately inertial regimes in the presence of significant surface tension effects, in which the shape of a freely deformable bubble is close to an oblate spheroid. \textcolor{black}{The bubble volume is $\mathcal{V}=\frac{4}{3}\pi R^3$, with $R$ denoting the equivalent radius.} Defining $b$ and $a$ as the lengths of the major and minor bubble semi-axes \textcolor{black}{(which are such that $ab^2=R^3$ in the case of a perfect spheroid),} the bubble geometrical aspect ratio is $\upchi=b/a$ (figure \ref{fig:problem-state}$(b)$). Given that the bubble motion predominantly occurs within the $(x,y)$ plane, the bubble orientation is characterized by two angles: the inclination $\alpha$, which is the angle from the vertical direction $\boldsymbol{e}_y$ to the minor axis, and the drift $\beta$, which is the angle from the minor axis to the bubble velocity $\boldsymbol{v}$, as illustrated in figure \ref{fig:problem-state}$(b)$. 
In what follows, the fluid velocity and pressure fields in the presence of the bubble are $\boldsymbol{u}$ and $p$, respectively, and $\boldsymbol\omega=\nabla\times\boldsymbol{u}$ denotes the vorticity.

The fluid and bubble motions are governed by the incompressible Navier-Stokes equations, together with the no-slip boundary condition  at the wall, $\boldsymbol{u}(0,y,z)=\boldsymbol{0}$. The problem is \textit{a priori} characterized by five independent dimensionless parameters, namely the Galilei number $Ga$, the Bond number $Bo$, the initial separation distance $X_0=x_0/R$, the density ratio $\rho_g / \rho_l$ and the viscosity ratio $\mu_g / \mu_l$, where subscripts $g$ and $l$ refer to the gas and the liquid, respectively. We set the density and viscosity ratios to $10^{-3}$ and $10^{-2}$, respectively, to mimic the situation of a gas bubble rising in a low-viscosity liquid. These two ratios are not expected to have any significant influence on the flow dynamics as long as they remain very small. The Galilei and Bond numbers are respectively defined as
	\begin{equation}
	Ga=\frac{\rho_l g^{1/2} R^{3/2}}{\mu_l}, \quad
	Bo=\frac{\rho_l g R^2}{\gamma}\,,
	\label{eq:ga-bo}
	\end{equation}
with $\gamma$ denoting the surface tension. The instantaneous bubble Reynolds number depends on $Ga$, $Bo$ and $X_0$ and is defined as
	\begin{equation}
	\Rey=\frac{2\rho_l  ||\boldsymbol{v}|| R}{\mu_l}\,,
	\label{eq:re}
	\end{equation}
with $||\boldsymbol{v}||$ the norm of the instantaneous velocity of the bubble centroid.
In what follows, we consider the parameter range $10\leq Ga \leq 30$ and $0.01\leq Bo \leq 1$ in which, with just one exception, an isolated bubble rising in an unbounded flow domain follows a vertical path. Note that, as far as we are aware, the only two previous numerical investigations of the same problem involving deformable bubbles focused on a single value of the Bond number, namely $Bo=4$ \citep{2020_Zhang} or $Bo=0.5$ \citep{2022_Yan}. Hence, we believe that results of fully-resolved simulations exploring the fate of nearly-spherical bubbles have not been reported so far. 

\textcolor{black}{
In experimental investigations under a given gravitational environment, the selected gas-liquid system may be characterized through the Morton number $Mo = g \mu_l^4 / (\rho_l \gamma^3) = Bo^3 / Ga^4$. The ranges of $Bo$ and $Ga$ considered in this work correspond to Morton numbers ranging from $1.2 \times 10^{-12}$ to $1.0 \times 10^{-4}$. 
To connect the $(Bo,Ga)$ sets discussed later with actual gas-liquid systems, table \ref{tab:mo} summarizes the physical properties of seven fluids with $Mo$ ranging from approximately $1\times10^{-12}$ to $1\times10^{-5}$, along with the corresponding Bond number and bubble size at the two extremities of the considered range of Galilei number, $Ga = 10$ and $Ga = 30$. 
}

\begin{table}
\centering
\begin{tabular}{lllllll}
Liquid              & $\rho_l~(\text{kg/m}^{3})$ & $\mu_l~(\text{mPa~s})$ & $\gamma~(\text{mN/m})$ & $Mo=\frac{g\mu_l^4}{\rho_l \gamma^3}$ & \multicolumn{2}{l}{$~~~~~~~~Bo\, (R\,\, \text{in mm})$} \\
                   &   &   &   &                  & $Ga=10~~~$          & $Ga=30$          \\\cline{6-7}
Iron at $1550^\circ\,$C               & 7010 & 6.30 & 1200 & $1.27\times10^{-12}$ & 0.002 (0.20)               & 0.010 (0.42)              \\
Water 		   & 1000 & 1.00 & 72.8 & $2.54\times10^{-11}$ & 0.006 (0.22)                & 0.027 (0.45)              \\
DMS-T00             & 761 & 0.49 & 15.9 & $1.8\times10^{-10}$ & 0.012 (0.16)                & 0.053 (0.33)                \\
DMS-T01             & 818 & 0.88 & 16.9 & $1.5\times10^{-9}$ & 0.025 (0.23)               & 0.107 (0.48)               \\
DMS-T02             & 873 & 1.75 & 18.7 & $1.6\times10^{-8}$ & 0.054 (0.35)                & 0.235 (0.72)                \\
DMS-T05             & 918 & 4.59 & 19.7 & $6.2\times10^{-7}$ & 0.184 (0.64)                & 0.795 (1.32)                \\
DMS-T11             & 935 & 9.35 & 20.1 & $9.9\times10^{-6}$ & 0.463 (1.01)                & 2.002 (2.10)                \\
\end{tabular}
\caption{\textcolor{black}{Physical properties of some fluids with $Mo$ ranging from $\approx1\times10^{-12}$ to $\approx1\times10^{-5}$. Except for iron, all fluid properties are taken at a temperature of $20^\circ\,$C. The last four columns on the right show, for $Ga = 10$ and $30$, the corresponding Bond number and, in parentheses, the equivalent bubble radius, $R$, in mm. } }
\label{tab:mo}
\end{table}

\subsection{Numerical framework and computational aspects}
The three-dimensional flow field is solved numerically using the open-source flow solver \emph{Basilisk} developed by Popinet \citep{2009_Popinet, 2015_Popinet}. This solver employs the one-fluid approach together with the \textcolor{black}{geometric volume of fluid (VOF) method} to track the bubble interface. The volume function $C({\bf{x}}, t)$, with $C=1$ within the bubble and $C=0$ in the liquid, determines whether gas or liquid is present at time $t$ at a given point $\bf{x}$ of the domain. The local density and viscosity of the fluid medium are approximated using the averaging rules
\refstepcounter{equation}
$$
\rho({\bf{x}}, t)=C({\bf{x}}, t) \rho_g + (1-C({\bf{x}}, t))\rho_l\,, \quad
\mu({\bf{x}}, t)=\frac{\mu_g\mu_l}{C({\bf{x}}, t)\mu_l + (1-C({\bf{x}}, t))\mu_g}
\eqno{(\theequation{\mathit{a},\mathit{b}})}
\label{eq:vof}
$$
A harmonic averaging rather than the more popular arithmetic averaging is used for the viscosity, as the latter is expected to somewhat over-estimate viscous effects. A comparison of results obtained with the two averaging rules is provided in appendix \ref{sec:appA}. Differences are found to be marginal (see figure \ref{fig:pre-test-mu}). \\
\indent The computational domain is a cubic box with an edge length $L = 240R$. A no-slip and no-penetration condition is applied at the left wall ($x = 0$) and at the top ($y = y_{\max}$) and bottom ($y = y_{\min}$) surfaces, while a free-slip condition is applied on the remaining boundaries. \textcolor{black}{To mimic a hydrophilic condition at the wall, we also impose $C=0$ at $x = 0$, thus enforcing the presence of a thin liquid film of thickness $\Delta_f$, with $0 < \Delta_f \leq \Delta_{\min}$ ($\Delta_{\min}$ denoting the minimum grid size), that cannot be entirely drained during a possible wall-bubble collision.} 
Bubbles are initially spherical and are released with zero velocity at a vertical position located $15R$ above the bottom wall to minimize confinement effects possibly induced by this wall. 

Numerical aspects of the interfacial flow solver incorporated in \emph{Basilisk}, particularly details on the VOF technique and the computation of surface tension, have been documented in many previous works [see, e.g., \citet{2009_Popinet} and \citet{2021_Zhang}]. Here, we only detail the spatial discretization used in this work, since this aspect is crucial for fully resolving the flow within the boundary layers that develop at the bubble surface and at the wall. 
We use the adaptive mesh refinement technique (AMR) to locally refine the grid close to the interface and in regions of high velocity gradients, based on a wavelet decomposition of $C$ and $\boldsymbol{u}$, respectively \citep{2018_Hooft}. The thresholds on the estimated relative error for $C$ and $\boldsymbol{u}$ are $10^{-3}$ and $10^{-2}$, respectively  \citep{2018_Hooft}, while the minimum and maximum grid sizes are $\Delta_{\min}=L/2^{14} \approx R/68$ and $\Delta_{\max}=L/2^{6} \approx4R$, respectively. These settings ensure that: (i) the boundary layer around the bubble surface is resolved using approximately six grid points, even at the highest Reynolds number reached in the considered parameter space; and (ii) the far wake (starting approximately $10R$ downstream of the bubble) is also adequately resolved, since the corresponding cells have a size close to $R/17$. To properly resolve the flow in the bubble-wall gap when the bubble is close to the wall, $\overline\Delta_{\min}\equiv\Delta_{\min}/R$ is decreased to $\approx 1/136$ when $\overline\delta\equiv\delta/R \leq 0.15$. The adequacy of the grid resolution is confirmed through a grid-independence study detailed in appendix \ref{sec:appA}, in which $\overline\Delta_{\min}$ is refined down to $1/272$ irrespective of $\overline\delta$. Other tests with bubbles rising below a horizontal or inclined wall are presented in appendix \ref{sec:appB}.\\
\indent A preliminary series of computations in the range $10\leq Ga \leq 30$, $0.02\leq Bo \leq 1$ was carried out using a coarser grid with $\overline\Delta_{\min}=(L/R)/2^{13} \approx 1/34$, to gain some insight into the organization of the flow field. These simulations revealed that the flow remains always symmetric with respect to the vertical mid-plane $z=0$. This allowed us to consider only half of the computational domain in the rest of the computations, by imposing a symmetry condition on that plane. The computations whose results are discussed below were performed on the HPC cluster \emph{Hemera} at the Helmholtz-Zentrum Dresden Rossendorf. Most runs were executed on nodes equipped with two Intel\textsuperscript{\textregistered} Xeon\textsuperscript{\textregistered} Gold 6148 CPUs processors, each comprising 20 cores and running at 2.40 GHz. A typical run, covering a physical time $t_{fin}=80 (R/g)^{1/2}$, took approximately 50 days (e.g., the case $Ga = 30$, $Bo = 0.25$). Computations became more time-consuming in low-$Bo$ cases, owing to the time step constraint in the explicit scheme involved in the computation of the capillary force \citep{2009_Popinet}. For this reason, computations with $Bo < 0.2$ were performed on nodes equipped with two AMD 16-Core Epyc 7302 CPUs processors, running at a higher clock speed of 3.0 GHz. Despite these improved performances, obtaining results covering a sufficiently long period of time usually required around 90 days (e.g., the case $Ga = 25$, $Bo = 0.05$). The most challenging case was the grid convergence study at $Ga = 21.9$ and $Bo = 0.073$ reported in appendix \ref{sec:appA}, where the grid was refined down to $\overline\Delta_{\min} = 1/272$. This refinement caused the typical number of grid cells in the second half of the run to increase from 2.8 million (when $\overline\Delta_{\min} = 1/136$) to about 8.1 million (when $\overline\Delta_{\min} = 1/272$), even though only half of the computational domain was considered. \textcolor{black}{Moreover, the time step was reduced from $6.8\times10^{-5}(R/g)^{1/2}$ to $2.4\times10^{-5}(R/g)^{1/2}$, owing to the constraint arising from capillary effects.} To tackle the memory issue inherent to this case, the run was executed on two AMD 64-Core Epyc 7713 CPUs processors running at 2.0 GHz. It took approximately half a year to obtain the evolution of the flow up to a physical time $t_{fin}=30 (R/g)^{1/2}$\textcolor{black}{, corresponding to approximately $1.25\times10^6$ time steps.}\\
\begin{table}
\centering
\renewcommand{\arraystretch}{1.5} 
\begin{tabular}{llccclccclcllllccc}
                     &  & \multicolumn{3}{c}{Length} &  & \multicolumn{3}{c}{Velocity} &  & \multicolumn{4}{c}{Time} &  & \multicolumn{3}{c}{Vorticity} \\ \cline{3-5} \cline{7-9} \cline{11-14} \cline{16-18} 
Dimensional variable     &  & $x_b$      & $y_b$      & $z_b$     &  & $v_x$       & $v_y$      & $v_z$      &  & \multicolumn{4}{c}{$t$}    &  & $\omega_x$       & $\omega_y$       & $\omega_z$      \\
Dimensionless variable   &  & $X_b$      & $Y_b$      & $Z_b$     &  & $V_x$       & $V_y$      & $V_z$      &  & \multicolumn{4}{c}{$T$}    &  & $\overline{\omega}_x$       & $\overline{\omega}_y$       & $\overline{\omega}_z$      \\
Characteristic scale &  & \multicolumn{3}{c}{$R$}      &  & \multicolumn{3}{c}{$\sqrt{gR}$}        &  & \multicolumn{4}{c}{$\sqrt{R/g}$}    &  & \multicolumn{3}{c}{$\sqrt{g/R}$}         
\end{tabular}
\caption{Correspondence between the dimensional and non-dimensional variables characterizing the bubble motion. Specifically, $v_i=\boldsymbol{v}\cdot\boldsymbol{e}_i$ and $\omega_i=\boldsymbol{\omega}\cdot\boldsymbol{e}_i$ represent the $i$-th component of velocity and vorticity, respectively.}
\label{tab:norm_var}
\end{table}
\indent In the following sections, we make an extensive use of dimensionless flow parameters to describe the bubble motion. Table \ref{tab:norm_var} summarizes the main dimensionless variables, alongside with their dimensional counterparts and the characteristic scales used for normalisation.  In particular, bubble positions, velocities and times will systematically be normalised by $R$, $(gR)^{1/2}$, and $(R/g)^{1/2}$, respectively. Unless stated otherwise, the dimensionless initial distance from the bubble to the wall is set to $X_0=2$, corresponding to an initial gap $\overline\delta|_{t = 0}= 1$. Effects of $X_0$ on the subsequent trajectory of the bubble are examined in appendix \ref{sec:appC}.  

\section{Path, wake and bubble dynamics}

\label{sec:path_wake}
	\subsection{Overview of the results}
	\label{sec:result_sum}

Three distinct types of near-wall bubble motions were observed in the series of computations we carried out.

\begin{itemize}
    \item \textit{Migration away from the wall}. In this scenario, the bubble continuously migrates away from the wall as soon as its wake is fully developed. This migration may or may not be accompanied by the development of path instability in the wall-normal plane.
    \item \textit{Periodic near-wall bouncing}. In such cases, the bubble bounces repeatedly, with a fixed amplitude and frequency, with or without `direct collisions' on the wall according to the definition detailed below.
    \item \textit{Damped near-wall bouncing}. In this scenario, the bubble first bounces several times very close to the wall, but the amplitude of the bounces decreases over time, and the bubble eventually stabilizes at a certain distance from the wall.
\end{itemize}

\begin{figure}
\vspace{5mm}
    \centerline{\includegraphics[scale=0.65]{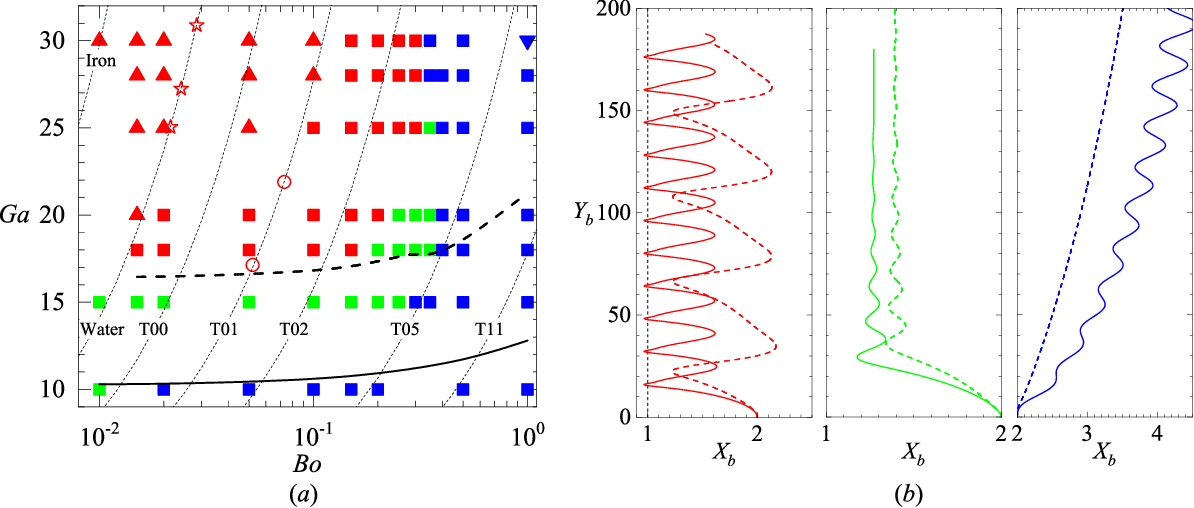}}
  \caption{Different types of bubble paths observed in the simulations. ($a$): Phase diagram in the ($Bo$, $Ga$) plane; ($b$): Typical trajectories illustrating the three distinct families of motion. \textcolor{red}{$\blacktriangle$} and \textcolor{red}{$\blacksquare$} in ($a$), \textcolor{red}{\full} and \textcolor{red}{$\dashed$} in ($b$): periodic bouncing cases with and without `direct' wall-bubble collisions, respectively; \textcolor{blue}{$\blacktriangledown$} and \textcolor{blue}{$\blacksquare$} in ($a$), \textcolor{blue}{\full} and \textcolor{blue}{$\dashed$} in ($b$): migration away from the wall with and without the development of a path instability, respectively; 
  \textcolor{green}{$\blacksquare$} in ($a$), \textcolor{green}{\full} and \textcolor{green}{$\dashed$} in ($b$): damped bouncing cases. In ($a$), open red stars and circles 
  refer to the experimentally observed periodic bouncing configurations of \citet{2002_devries} and TM, respectively;  \textcolor{black}{\full},\, \textcolor{black} {\dashed}: iso-Reynolds numbers lines corresponding to the critical values $\Rey_1=35$ and $\Rey_2=65$, respectively; \textcolor{black}{thin dashed lines are the iso$-Mo$ lines corresponding to different liquids, with iron and water at the very left, then silicone oils T0-T11 of increasing viscosity from left to right (see table~\ref{tab:mo} for the corresponding physical properties).} In ($b$), lines \textcolor{red}{\full},\, \textcolor{red} {\dashed},\, \textcolor{green}{\full},\, \textcolor{green}{\dashed},\, \textcolor{blue}{\full},\, and \textcolor{blue}{\dashed} correspond to parameter combinations $(Bo, Ga)=(0.05,25)$, $(0.25,30)$, $(0.05,15)$, $(0.25,20)$, $(1,30)$, and $(0.5,15)$, respectively.
}
    \label{fig:traj_sum}
\end{figure}

Figure \ref{fig:traj_sum} displays the phase diagram and typical trajectories associated with these three families of motions. The two trajectories in the left panel of figure~\ref{fig:traj_sum}$(b)$ correspond to the periodic near-wall bouncing regime. The path denoted with a solid red line is observed with a nearly-spherical bubble ($\upchi\approx1.05$; see figure~\ref{fig:regular_bounce_motion1}$(b1, b2)$ below) which repeatedly collides with the wall (collisions arise at positions where $\min(X_b)<1$). The second path (dashed red line), corresponds to a moderately deformed bubble ($\upchi\approx1.5$; see figure~\ref{fig:regular_bounce_motion2}$(b)$ below) which does not collide with the wall throughout its ascent but bounces periodically close to it. Note that the amplitude of the lateral motion  is significantly larger in this second case. The middle panel in figure~\ref{fig:traj_sum}$(b)$ displays the trajectories of two bubbles exhibiting a damped near-wall bouncing dynamics. The bubble that follows the path denoted with a green solid line is nearly spherical ($\upchi\approx1.05$), while the one that rises along the green dashed path is moderately deformed ($\upchi\approx1.3$); see figure~\ref{fig:damp_bounce_motion}$(b)$. Finally, the third panel in figure~\ref{fig:traj_sum}$(b)$ illustrates the trajectories of two bubbles that migrate away from the wall during their ascent. The path denoted with a solid blue line is that of a bubble with $(Bo, Ga)=(1,30)$ experiencing path instability; this bubble would stay in the zigzagging regime if rising in an unbounded domain~\citep{2016_Cano-Lozano}.\\
\indent In the moderate Reynolds number regime, the transverse force acting on a spherical bubble rising some distance from a vertical wall reverses from repulsive to attractive beyond a critical Reynolds number $\Rey_1\approx35$ (TM; \cite{2015_Sugioka, 2020_Shi, 2024_Shi}). The iso-Reynolds number line $\Rey=35$ drawn in figure \ref{fig:traj_sum}$(a)$  
confirms that, regardless of their Bond number, bubbles with $\Rey\lesssim35$ migrate away from the wall (the case $(Bo, Ga) = (0.01, 10)$ in which the bubble is found to experience damped bounces is marginal, since the corresponding Reynolds number is $\Rey= 34.5$, very close to $\Rey_1$). Conversely, the transverse motion of bubbles with $\Rey > 35$ depends on their oblateness. Specifically, nearly-spherical bubbles with Bond numbers in the range $0 < Bo < 0.25$ are attracted to the wall, as expected. Their path may exhibit regular or damped bounces, depending on a second critical Reynolds number to be discussed below. Bubbles with $Bo > 0.25$ exhibit a significant oblateness and tend to depart from the wall, even for $\Rey>35$. Since the repulsive mechanism results from the interaction of the bubble wake with the wall, the stronger the wake (hence the vorticity at the bubble surface), the larger the repulsive force. More specifically, physical arguments developed by \cite{2002_Takemura} and TM indicate that this force varies approximately as the square of the maximum vorticity at the bubble surface. This vorticity being directly proportional to the curvature of the gas-liquid interface \citep{1967_Batchelor}, distorted bubbles are expected to keep on being repelled from the wall at higher Reynolds number than nearly-spherical bubbles. 
At a fixed $Ga$ such that $\Rey(Ga)\gtrsim35$, this influence of the bubble oblateness results in a change of sign of the transverse force flipping from attractive to repulsive beyond a certain critical Bond number. According to figure \ref{fig:traj_sum}$(a)$, this critical $Bo$ is close to $0.35$ when $Ga\gtrsim 18$.\\
\indent Periodic near-wall bouncing motion was observed experimentally at moderate-to-high Reynolds number by \citet{2002_devries} and TM. The corresponding data are shown in figure \ref{fig:traj_sum}$(a)$ (open stars and circles). They lie well within the region where the same type of behaviour is detected in the present work. With nearly-spherical bubbles, TM noticed that a repeatable near-wall bouncing motion takes place beyond a second critical Reynolds number $\Rey_2\approx65$ (dashed line in figure \ref{fig:traj_sum}$(a)$).  Present results agree well with this criterion, since all cases in which the bubble is observed to bounce repeatedly lie above this dashed line as long as $Bo\leq0.1$. For larger Bond numbers, present results indicate that $\Rey_2$ increases with increasing $Bo$, owing to the increased amount of vorticity generated at the bubble surface for the reason mentioned above.  \\
 \indent In the intermediate range $\Rey_1 < \Rey < \Rey_2$, bubbles with low-to-moderate deformation, say those with $Bo\leq0.25$, are seen to perform a damped bouncing motion; two examples are shown in figure \ref{fig:traj_sum}$(b)$. For instance, a bubble with $Bo=0.01, Ga=15$ {\color{black}{(not shown in figure  \ref{fig:traj_sum}$(b)$)}} finally stabilizes at a wall-normal position  $X_b=1.24$, and rises with a final Reynolds number $\Rey\approx57.6$. These findings are in good agreement with the predictions of simulations performed with a fixed spherical bubble \citep{2024_Shi}, in which the transverse dimensionless position at which the overall lateral force vanishes was found to be close $1.25$ for $\Rey =55$. 
\begin{figure}
\vspace{5mm}
    \centering
    \includegraphics[scale=0.8]{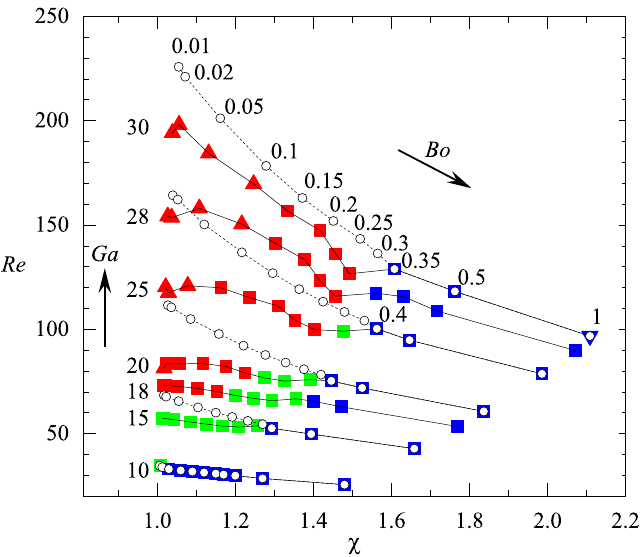}
    \caption{The three regimes of bubble-wall interaction in the ($\Rey$, $\upchi$) plane. \textcolor{black}{Solid symbols and open circles denote values of ($\Rey$, $\upchi$) obtained under wall-bounded and unbounded conditions, respectively.} \textcolor{black}{The color and shape codes of the solid symbols are identical to those of figure \ref{fig:traj_sum}}. 
    Values of $\Rey$ and $\upchi$ \textcolor{black}{in the presence of the wall} are based on averages taken over a single period of bounce in the periodic bouncing regime, and on final values in the other two regimes.  \textcolor{black}{Solid and dashed lines denote iso-$Ga$ lines in the wall-bounded and unbounded configurations, respectively,} with $Ga$ increasing from $10$ to $30$ from bottom to top, and $Bo$ increasing from $0.01$ to $1$ from left to right.}
    \label{fig:chi-vs-re}
\end{figure}
\begin{figure}
    \centering
    \includegraphics[scale=0.65]{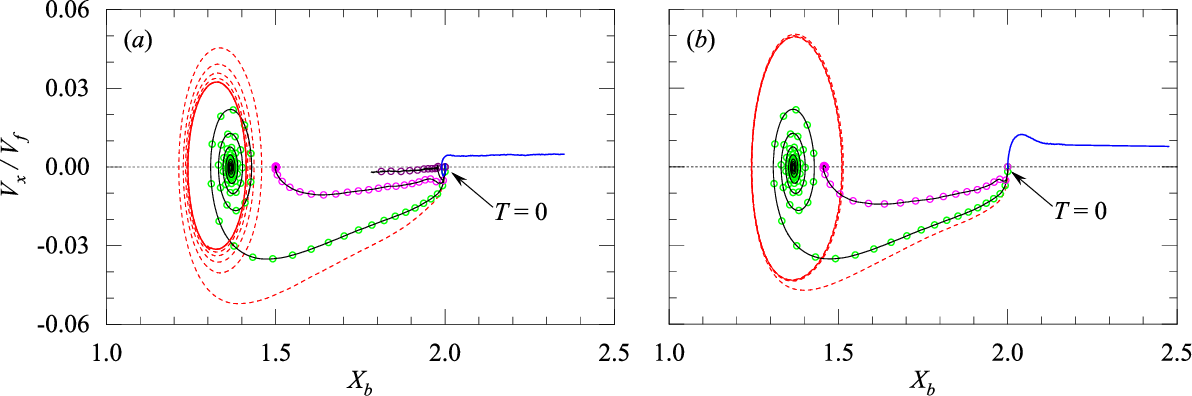}
    \caption{Variations of the bubble wall-normal velocity, $V_x$, normalised by the terminal velocity, $V_f$, as a function of the the wall distance, $X_b$ (in the regular bouncing configurations, $V_f$ is taken to be the mean rise speed, $V_m$). $(a)$ $\Rey\approx67$, with $\upchi=1.15$ ($\textcolor{red}{\dashed}$), $1.2$ (\textcolor{green}{$\circ$}), $1.29$ (\textcolor{magenta}{$\circ$}), $1.36$ (\textcolor{purple}{$\circ$}), $1.4$ ($\textcolor{blue}{\full}$); 
    $(b)$ $\upchi\approx1.2$, with $\Rey=79$ ($\textcolor{red}{\dashed}$), $68$ (\textcolor{green}{$\circ$}), $53$ (\textcolor{magenta}{$\circ$}), and $30$ ($\textcolor{blue}{\full}$). Data in $(a)$ correspond to cases with fixed $Ga=18$ and $Bo=$ 0.15, 0.2, 0.3, 0.35, and 0.4, respectively, while those in $(b)$ correspond to $(Bo,Ga)=$  (0.2,20), (0.2, 18), (0.25, 15), and (0.3,10), respectively. In each series, the initial position is $(X_b, V_x) = (2, 0)$.}
    \label{fig:vx_2_vy_damp}
\end{figure}
Figure \ref{fig:chi-vs-re} is the equivalent of figure \ref{fig:traj_sum}$(a)$ in the $(Re,\upchi)$ plane. \textcolor{black}{Values of the terminal Reynolds number and aspect ratio for a given $(Ga,Bo)$ set in an unbounded domain (open circles) are given as reference, to better appreciate the influence of the wall. Note that in the case of bubbles migrating away from the wall (blue symbols), there is no distinction between the resulting ($\Rey$, $\upchi$) with or without the wall, since in that regime bubbles no longer experience any wall effect in the final stage of their motion. In wall-bounded configurations,} the figure allows to determine directly the influence of the Reynolds number and bubble aspect ratio on the transition between the different regimes. \textcolor{black}{It is worth noting that, for a given $Ga$, the observed type of path does not change with the aspect ratio up to $\upchi\approx1.2$, underlining that $Ga$ (or $\Rey$) is the only significant control parameter of the system for nearly spherical bubbles.} Figure \ref{fig:chi-vs-re} also emphasizes the limited range of aspect ratios in which the periodic bouncing regime exists, from $\upchi\lesssim1.16$ for $Ga=18$ ($\Rey\approx70$) to $\upchi\lesssim1.5$ for $Ga=30$ ($\Rey\approx125$). The marked increase in the critical Reynolds number $\Rey_2$ with the bubble deformation is also noticeable, the damped bouncing regime being observed up to $\Rey\approx100$ with a bubble having $\upchi=1.48$ for $Ga=25$. 
Figure \ref{fig:vx_2_vy_damp} depicts the variations of the wall-normal velocity $V_x(T)$ (normalised by the terminal rise speed, $V_f$) with the lateral position $X_b(T)$ in the various scenarios at a fixed Reynolds number or a fixed aspect ratio. Setting $\Rey\approx67$ (figure \ref{fig:vx_2_vy_damp}$(a)$), regular bounces are prominent up to $\upchi\approx 1.15$. The damped bouncing regime emerges for larger oblatenesses. Bounces are underdamped up to $\upchi\approx1.25$ and become overdamped with more oblate bubbles, the bubble then reaching its final equilibrium position without any overshoot. For $\upchi\gtrsim 1.36$, the bouncing motion ceases, and bubbles are consistently repelled from the wall. Setting $\upchi\approx1.2$ (figure \ref{fig:vx_2_vy_damp}$(b)$), a similar transition takes place with decreasing $\Rey$. Here, regular bouncing is observed at $\Rey=79$. As $\Rey$ decreases, the motion transitions to underdamping at $\Rey=68$, then to overdamping at $\Rey=53$, and the uniform migration away from the wall is eventually observed at $\Rey=30$. The qualitatively similar evolutions observed in figure \ref{fig:vx_2_vy_damp} by increasing $\upchi$ at a given $\Rey$ or decreasing $\Rey$ at a given $\upchi$ may be interpreted through the prism of the relative magnitudes of the irrotational and vortical interaction mechanisms. Clearly, the irrotational component weakens as $\Rey$ decreases, due to the increasing magnitude of viscous effects. Similarly, increasing $\upchi$ at a given $\Rey$ increases the amount of vorticity produced at the bubble surface \citep{2007_Magnaudet}, thus strengthening vortical effects. 

\subsection{Migration away from the wall}
\label{sec:depart_from_wall}
\begin{figure}
\vspace{5mm}
    \centering
    \includegraphics[scale=0.65]{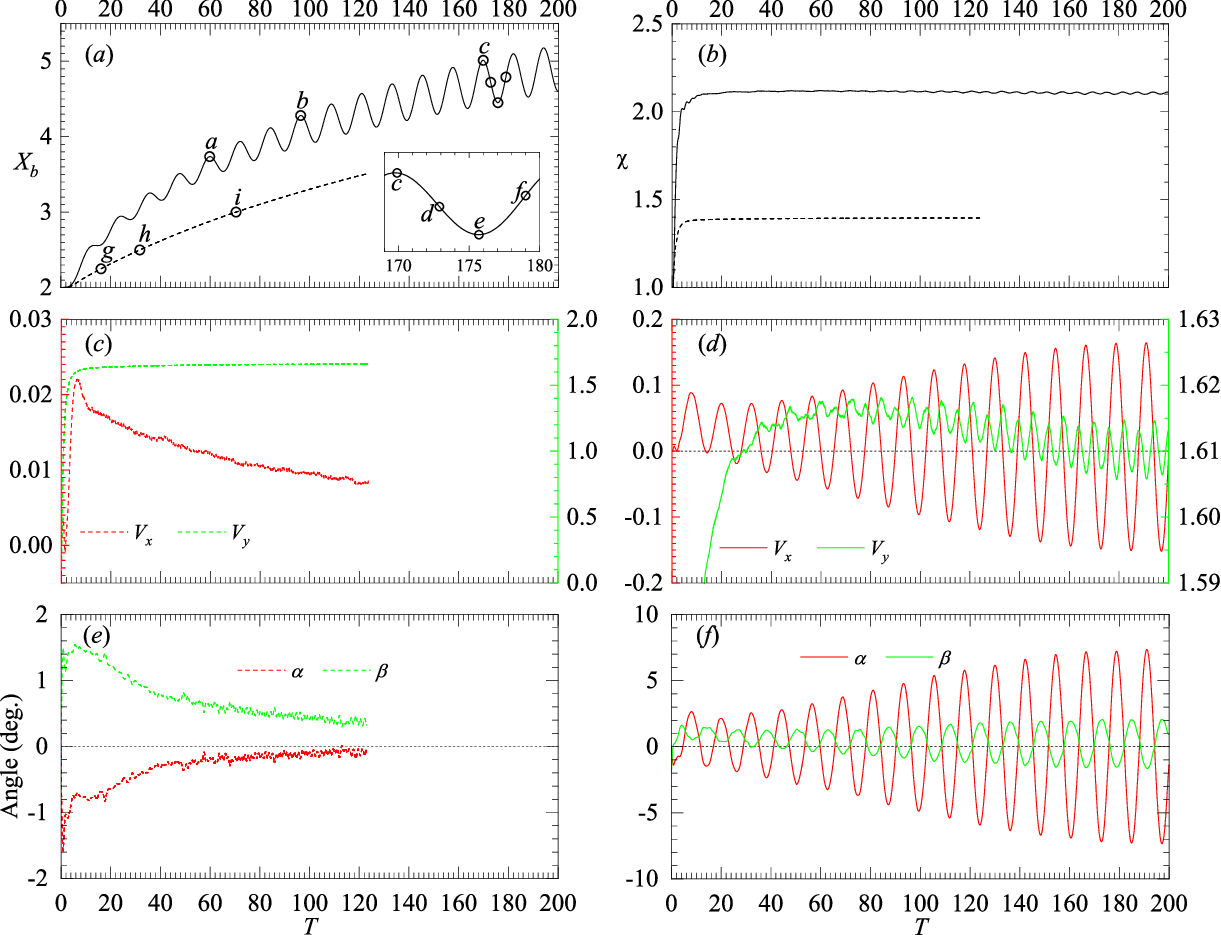}
    \caption{Evolution of various non-dimensional characteristics during the lateral migration of a bubble with $(Bo, Ga) = (1, 30)$ (solid lines) and $(0.5, 15)$ (dashed lines). ($a$): wall-normal position of the centroid; ($b$): aspect ratio; ($c-d$): components of the velocity of the bubble centroid; ($e-f$): inclination and drift angles. In panels $(c-d)$, the left and right axes refer to the horizontal ($V_x$) and vertical ($V_y$) velocity components, respectively.}
    \label{fig:depart_motion}
\end{figure}

In this section and those that follow, we examine in more detail the results obtained in each of the three regimes identified above. \\
\indent Figure \ref{fig:depart_motion} depicts the evolution of various characteristics of the bubble motion for the two sets of parameters $(Bo, Ga) = (1, 30)$ (solid lines) and $(0.5, 15)$ (dashed lines). In both cases, the bubble migrates away from the wall, as the evolution of the horizontal position of its centroid in figure \ref{fig:depart_motion}$(a)$ confirms. With $(Bo, Ga) = (0.5, 15)$, figures \ref{fig:depart_motion}$(b-c)$ indicate that, beyond $T\approx100$, the bubble achieves an equilibrium aspect ratio $\upchi=1.4$ and a rise Reynolds number \textcolor{black}{$\Rey=2Ga\,V_y\approx50$, since the terminal rise speed is close to $1.65$}. In contrast, path instability takes place in the case $(Bo, Ga) = (1, 30)$, leading to the emergence of a planar zigzagging path. This is not unlikely, since in an unbounded fluid path instability for a bubble with $Bo=1$ sets in at $Ga\approx29.65$ \citep{2023_Bonnefis_b}.  According to panels $(d)$ and $(f)$, the wall-normal bubble velocity and the inclination angle increase until $T\approx180$, suggesting that path instability has virtually saturated at the end of the run.  In this late stage, the bubble aspect ratio and average Reynolds number are $\upchi\approx2.1$, $\Rey\approx96.7$ (figures \ref{fig:depart_motion}$(b, d)$). Panels $(e-f)$ show that in both cases the drift angle $\beta$ remains small throughout the bubble ascent, ensuring that the path and the minor axis of the bubble remain almost aligned throughout the ascent, i.e. the bubble moves essentially broadside on. For $(Bo, Ga) = (1, 30)$, $\beta$ oscillates around zero with a maximum of approximately $2^{\circ}$, in quantitative agreement with the value determined in an unbounded fluid \citep{2001_Ellingsen, 2006_Mougin}. Since the bubble rise speed reaches its maximum twice during a period of the zigzag, the oscillation frequency of $X_b$, $V_x$, $\alpha$, and $\beta$ is half that of $V_y$. 


\begin{figure}
\vspace{5mm}
    \centering
    \includegraphics[scale=0.6]{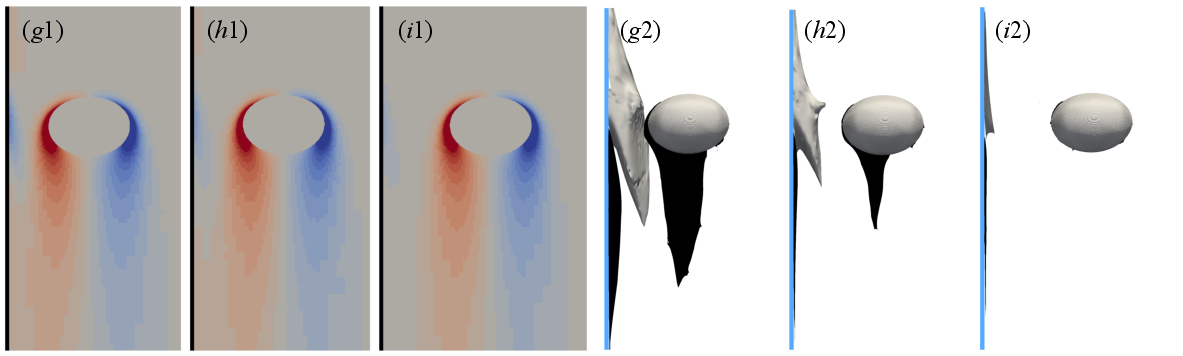}
    \caption{Evolution of the vortical structure past a bubble with $(Bo, Ga) = (0.5, 15)$ migrating away from the wall. Instants corresponding to panels $(g-i)$ are indicated by circles in figure \ref{fig:depart_motion}$(a)$. $(g1-i1)$: iso-contours of the normalised spanwise vorticity $\overline{\omega}_z$ in the symmetry plane $z=0$ (red and blue colours correspond to positive and negative $\overline{\omega}_z$, respectively); $(g2-i2)$: iso-surfaces of the normalised streamwise vorticity $\overline{\omega}_y=\pm 0.02$ in the half-space $z<0$ (grey and black threads correspond to positive and negative $\overline{\omega}_y$, respectively). In each panel, the wall lies on the left, represented by a vertical line. }
    \label{fig:depart_vor2}
\end{figure}

\begin{figure}
\vspace{2mm}
    \centering
    \includegraphics[scale=0.65]{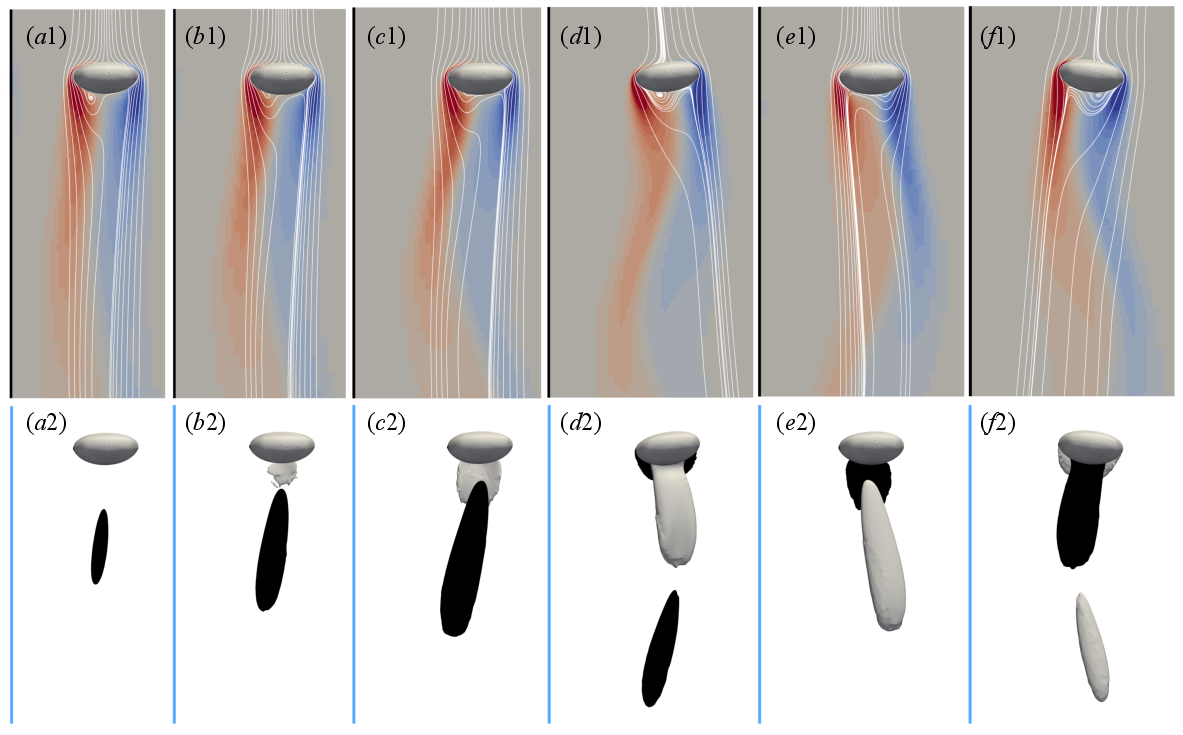}
    \caption{Same as figure \ref{fig:depart_vor2} for a bubble with $(Bo, Ga) = (1,30)$. 
 In the first row, the maximum of $|\overline{\omega}_z|$ is $2.0$; some streamlines computed in the bubble reference frame are displayed in the form of white lines; in the second row, the two iso-surfaces correspond to $\overline{\omega}_y=\pm 0.2$. 
    }
    \label{fig:depart_vor1}
\end{figure}

Figures \ref{fig:depart_vor2} and  \ref{fig:depart_vor1} reveal the vortical structure of the flow past the bubble in the above two cases at several instants of their rise, specified in figure \ref{fig:depart_motion}$(a)$. 
 Panels $(g1-i1)$ in figure \ref{fig:depart_vor2} suggest that the near wake of a bubble with $(Bo, Ga) = (0.5, 15)$ is almost axisymmetric, even when the gap is of the order of the bubble size. Nevertheless, owing to the interaction of the wake with the wall, a weak streamwise component exists in the vorticity field, both at the wall and at the back of the bubble. As panels $(g2-i2)$ highlight, this component decays rapidly as the separation increases. 
The situation is dramatically different for the bubble with $(Bo, Ga) = (1, 30)$. Here, the $\overline{\omega}_z$-distribution exhibits a marked asymmetry at all times and so do the streamlines, especially just at the back of the bubble (figure \ref{fig:depart_vor1}$(a1-f1)$). 
As soon as the lateral path oscillations have reached a significant amplitude, the two counter-rotating streamwise vortices characteristic of zigzagging bubbles become visible in the wake (figure \ref{fig:depart_vor1}$(c2-f2)$), with their orientation alternating every half-period of the zigzag. Note that no significant streamwise vorticity is present at the wall in these stages, suggesting that the wall plays only a marginal role in the bubble dynamics. Nevertheless, in the early stages, the wall dictates the orientation of the plane in which the bubble oscillates, making path instability arise through an imperfect bifurcation. 

\subsection{Periodic near-wall bouncing}
\label{sec:regular_bounce}
\begin{figure}
\vspace{5mm}
    \centering
    \includegraphics[scale=0.65]{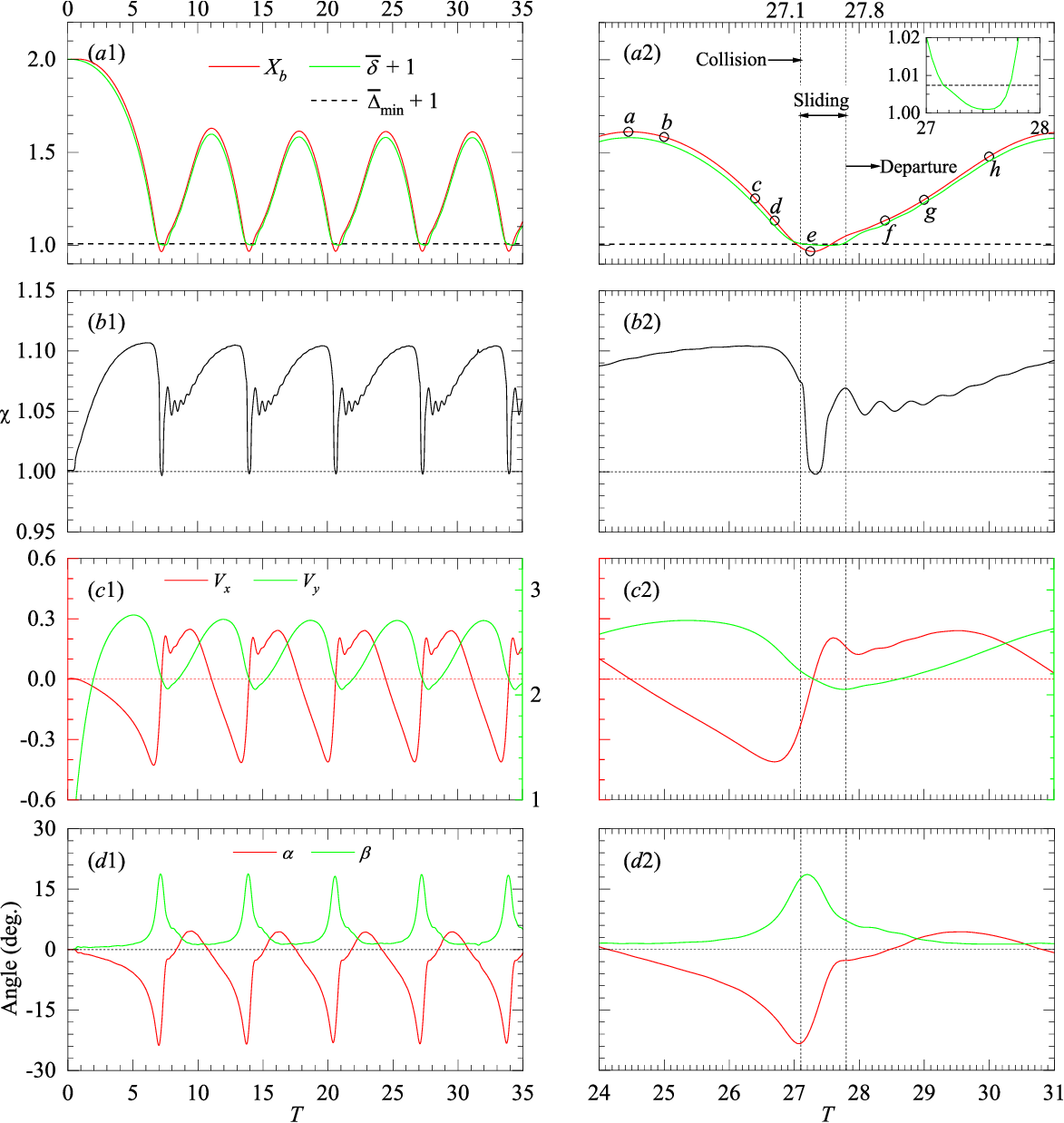}
    \caption{Evolution of various characteristics of the bubble and path during a periodic series of bounces for $(Bo, Ga) = (0.05, 25)$. Plots in the right column provide details of the evolution shown in the left column over a single bouncing period. ($a1-a2$): wall-normal position of the centroid (red line) and  gap thickness (green line); ($b1-b2$): aspect ratio;  ($c1-c2$): velocity components of the bubble centroid; ($d1-d2$): inclination and drift angles. In panels $(c1-c2)$, the left and right axes refer to $V_x$ and $V_y$, respectively.}
    \label{fig:regular_bounce_motion1}
\end{figure}
\begin{figure}
\vspace{5mm}
    \centering
    \includegraphics[scale=0.65]{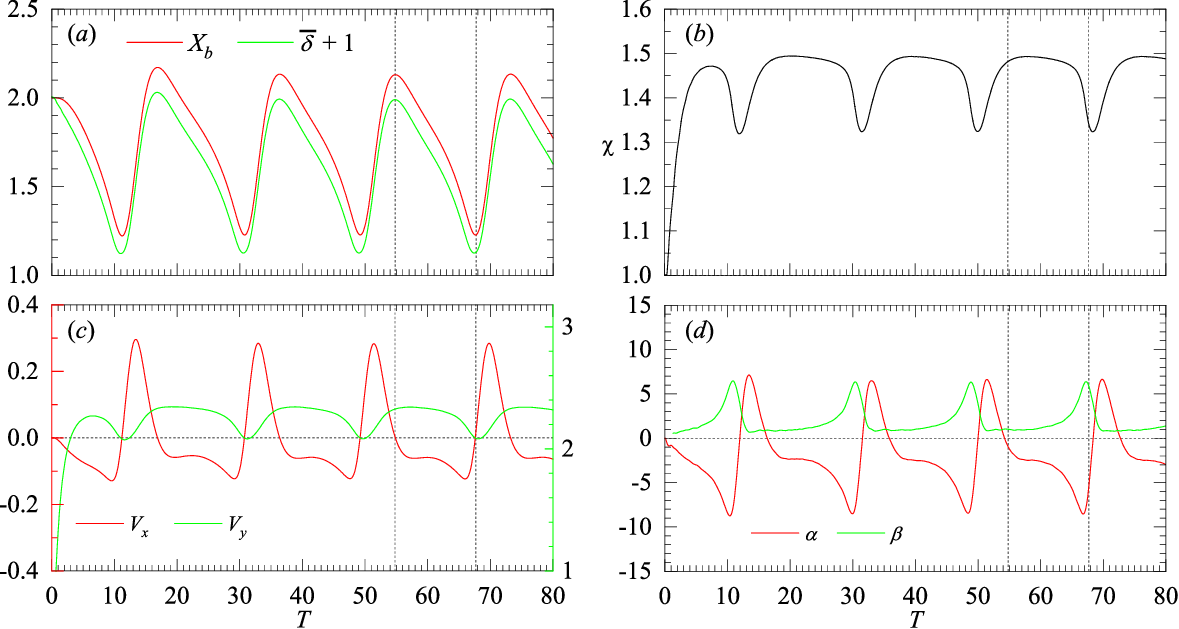}
    \caption{Same as figure \ref{fig:regular_bounce_motion1} for $(Bo, Ga) = (0.25, 30)$.}
    \label{fig:regular_bounce_motion2}
\end{figure}

Figures \ref{fig:regular_bounce_motion1} and \ref{fig:regular_bounce_motion2} illustrate the evolution of various characteristics of the bubble and its path in the case of bubbles with $(Bo, Ga) = (0.05, 25)$ and $(0.25, 30)$, both of which experience a periodic series of bounces. The key difference between these two configurations is the occurrence of direct wall-bubble collisions in the former case. Collision events can be identified from the temporal evolution of the dimensionless gap, $\overline\delta(T)$. As figure \ref{fig:regular_bounce_motion1}$(a2)$ shows, $\overline\delta(T)$ decreases to about half the minimum grid size at regular time intervals, first at $T\approx7$.  We define this configuration, as well as all those in which the minimum of $\overline\delta$ is smaller than or equal to $\overline\Delta_\text{min}$, as a `direct collision' between the bubble and wall in the sense of the macroscopic description allowed by the simulations. In such cases, the flow within the very thin liquid film remaining along the wall is of course not properly resolved. \textcolor{black}{It is important to identify the phenomena that are not captured by the imposed resolution and how much they affect the corresponding predictions. In the approaching stage, the liquid film is squeezed by the displacement of the bubble surface, similar to what happens when a rigid particle approaches a wall at right angle \citep{1999_Zenit}. This results in an outward semi-Poiseuille flow in the gap, given the no-slip condition at the wall and the shear-free condition at the the gas-liquid interface. If the bubble remained perfectly spherical ($Bo=0$), the overpressure in the thinnest part of the gap would result in a repulsive viscous force proportional to $V_x$ and inversely proportional to $\overline\delta$ \citep{2019_Michelin}. This force diverges as $\overline\delta\rightarrow0$ and therefore reaches very large values when the gap becomes of the order of the critical distance ($\approx10\,$nm) at which non-hydrodynamic effects become dominant. However, finite-deformation effects prevent this extreme situation from occurring: as figure \ref{fig:mesh}$(c)$ shows for a configuration close to the present one ($Bo=0.073, Ga=21.9)$, the part of the bubble surface facing the wall flattens as the gap reduces, increasing dramatically the area over which the overpressure applies. For Bond numbers in the range $[0.01,0.1]$, this makes the global lubrication force acting  on the bubble reach large values for minimum gaps much thicker than the above critical distance, stopping the motion of such bubbles towards the wall when the minimum film thickness is in the micrometre range. The grid convergence study presented in appendix \ref{sec:appA} helps quantify the influence of under-resolution in situations of `direct collision'. The evolutions of $X_b(T)$ and $\overline\delta(T)$ in figures \ref{fig:regular_bounce_motion1}$(a1,a2)$ are similar to those obtained in this test case with $\overline\Delta_\text{min}=1/68$ (red lines in figures \ref{fig:pretest1}$(d-e)$). Convergence is achieved in the test case by refining the grid down to $\overline\Delta_\text{min}=1/272$ (blue line). The main macroscopic difference between the results provided by the two resolutions is that the maximum separation reached by the bubble, hence the amplitude of the oscillations, is increased by a few percent with the finest grid. This is no surprise, since the better capture of lubrication effects in the gap results in an increase in the repulsive force acting on the bubble. In contrast, no sizeable change is noticed in the frequency of the oscillations. Based on this test case, results obtained in `direct collision' configurations appear to be reliable, although they certainly somewhat underestimate the amplitude of the oscillations. In the present case, there is little doubt that `direct collision' would be avoided  if the grid in the gap were refined by a factor of four, down to $\overline\Delta_\text{min}=1/544$.}\\
\indent \textcolor{black}{Figure \ref{fig:regular_bounce_motion1}$(a2)$} covering the time window $T\in[24,31]$ allows a more detailed appreciation of the succession of events occurring during a bounce.  As $T$ increases from 27.1 to 27.8, the bubble is seen to slide along the wall, with the gap remaining such that $\overline\delta\leq\overline\Delta_\text{min}$. Immediately after the collision, the bubble undergoes significant shape oscillations. Specifically, the aspect ratio, which is close to $1.1$ before the collision, drops rapidly to unity at $T\approx27.3$. Then it rises to $1.07$ when the gap starts to re-increase ($T\approx27.8$). Subsequently, the bubble departs from the wall and a series of oscillations with a dimensionless radian frequency close to $4\pi\approx12.57$ takes place. This frequency is comparable with that associated with the `mode 2' shape oscillations of a nearly spherical bubble, namely $f_2=\sqrt{12}Bo^{-1/2}\approx15.49$ \citep{1932_Lamb}. These findings are in line with the experimental observations reported by \citet{2002_devries} (see figure 5 therein) in a much more inertial regime ($Bo=0.1, Ga=76.2$). \\
\indent The above shape oscillations are intimately connected with the evolution of the bubble velocity. As depicted in figure \ref{fig:regular_bounce_motion1}$(c2)$, starting from $T\approx27.0$, the (still negative) wall-normal velocity experiences a rapid increase, changing sign at $T=27.3$ and achieving a local peak at $T\approx27.6$, just before $\upchi$ reaches a local maximum. This is followed by a short but sharp decrease, then by a secondary increase, with $V_x$ reaching its next peak at $T\approx29.5$. This is the time by which shape oscillations almost vanish. During the sliding stage where $V_x$ reverses, the vertical velocity $V_y$ decreases continuously, reaching its minimum at $T=27.8$. The decrease (resp. increase) of the aspect ratio as the bubble gets very close to (resp. departs from) the wall plays a role in the bouncing dynamics. Indeed, decreasing $\upchi$ increases the frontal area involved in the transverse motion (this area goes like $\upchi^{-1/3}$). This acts to increase the drag associated with the transverse motion, as well as the amount of liquid displaced by the bubble in that motion, i.e., its transverse virtual mass. Both aspects cooperate to slow down the transverse motion towards the wall. The arguments reverse after the bounce, indicating that the flattening of the bubble associated with the subsequent increase of $\upchi$ somewhat helps the bubble depart from the wall. Variations of the instantaneous rise speed follow directly those of the lateral bubble position. This is a direct consequence of the no-slip condition that forces $V_y(T)$  to decrease when the bubble gets very close to the wall, and to re-increase when it moves away from it. These variations have in turn a direct influence on the bubble shape. Indeed, when the Reynolds number is large, the oblateness of slightly non-spherical bubbles is known to be linearly proportional to the Weber number, $We=Bo(V_x^2+V_y^2)$ \citep{1965_Moore}. Based on figure \ref{fig:regular_bounce_motion1}$(c1)$, \textcolor{black}{the Weber number is found to vary by more than $40\%$ during a period of bounce, reducing from $0.37$ to $0.21$ as the bubble approaches the wall}, which directly translates into significant variations of $\upchi$ in between the two extreme bubble positions. In summary, it appears that time variations of the bubble shape result partly from variations of the rise speed imposed by the no-slip condition, and partly from the alternating transverse motion inherent to the bounce dynamics. \\
\indent The evolution of the bubble and path characteristics in the case $(Bo, Ga) = (0.25, 30)$ are displayed in figure \ref{fig:regular_bounce_motion2}. Marked differences with the previous observations may be noticed. The most obvious of them is that the minimum gap thickness is now close to $0.12$, indicating that the wall remains covered by a liquid film with a significant thickness all along the sequence of bounces. The aspect ratio, wall-normal velocity and inclination angles are seen to experience much more regular variations than in the previous case. In particular, the aspect ratio is found to oscillate smoothly from a minimum value \textcolor{black}{$\upchi\approx1.32$} reached slightly after the wall-normal velocity returns to positive, to a maximum value $\upchi\approx1.49$ when the bubble is `far' from the wall. These oscillations are far from sinusoidal, $\upchi$ keeping values close to its maximum for a long time and experiencing sharp variations only during short periods of time. Since the two cases essentially differ by the value of the Bond number (the two $Ga$ are close), the above differences underline the decisive influence of bubble deformability on the bounce dynamics. Nevertheless, despite these differences, the evolutions of the flow parameters in the two cases also share a number of generic features. 
First, the rise speed $V_y$ and the transverse velocity $V_x$ oscillate with the same dominant frequency. 
This stems from the retarding effect imposed by the wall on the rise speed \citep{2002_Takemura, 2015_Sugioka, 2020_Shi}. As mentioned above, this effect makes $V_y$ follow the variations of $X_b-1$, which in turn 
largely enslaves the evolution of the aspect ratio to that of $V_y$ (figures \ref{fig:regular_bounce_motion1}$(b1, c1)$ and \ref{fig:regular_bounce_motion2}$(b, c)$).
Second, figures~\ref{fig:regular_bounce_motion1}$(c1-d1)$ and \ref{fig:regular_bounce_motion2}$(c-d)$ show that the inclination angle $\alpha$ and the wall-normal velocity follow closely similar evolutions. In particular, both quantities reach their extrema simultaneously: the bubble axis bends toward the wall when $V_x<0$ (this is the configuration observed over $65\%$ to $75\%$ of the period between two bounces), and towards the fluid interior when $V_x>0$. 
Additionally, the drift angle $\beta$ remains consistently positive throughout the bubble ascent, irrespective of the sign of the wall-normal velocity. 

\begin{figure}
\vspace{5mm}
    \centering
    \includegraphics[scale=0.65]{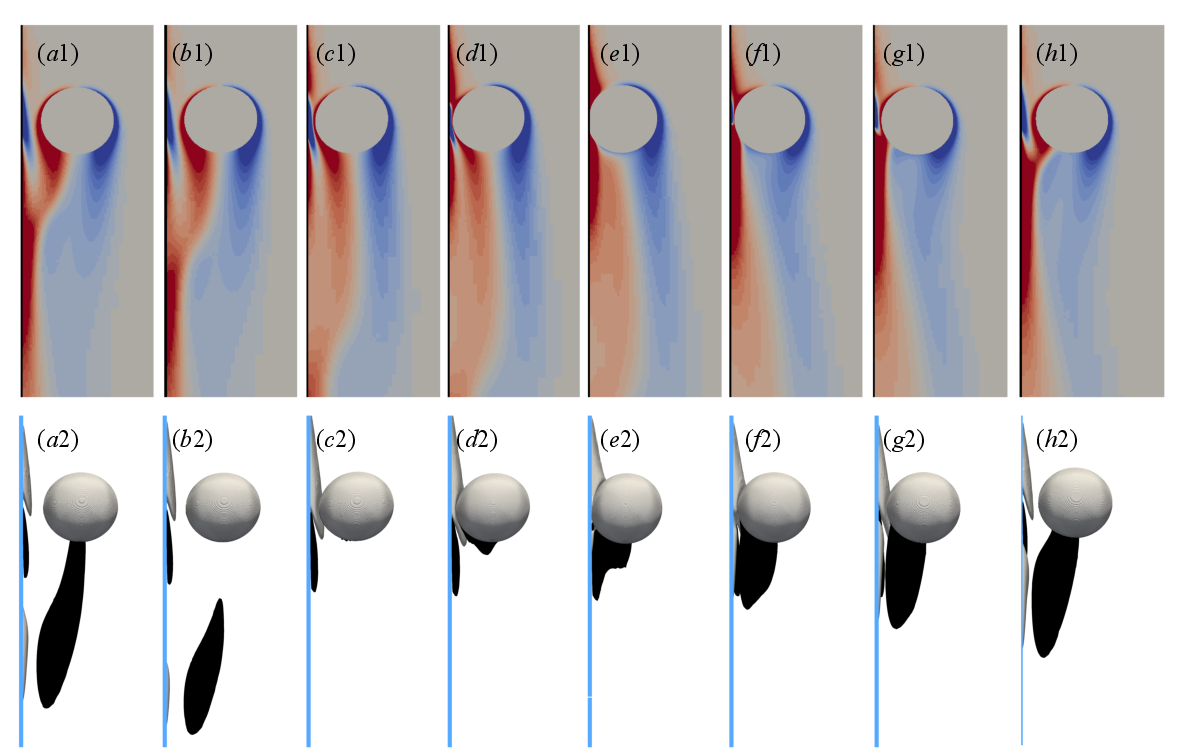}
    \caption{Evolution of the vortical structure past a bubble with $(Bo, Ga) = (0.05, 25)$ over one period of bouncing. From left to right in each row, the snapshots correspond to the successive bubble positions marked with circles in figure \ref{fig:regular_bounce_motion1}$(a2)$. Top row: iso-contours of the normalised spanwise vorticity $\overline{\omega}_z$ in the symmetry plane $z=0$ (red and blue contours indicate positive and negative $\overline{\omega}_z$, respectively, with a maximum magnitude of $2.0$); bottom row: iso-surfaces $\overline{\omega}_y= -0.5$ (black) and $+0.5$ (grey) of the streamwise vorticity in the half-space $z<0$. In each panel, the wall lies on the left and is represented by a vertical line.}
    \label{fig:regular_bounce_vor1}
\end{figure}

Figure \ref{fig:regular_bounce_vor1} displays the evolution of the spatial distribution of the spanwise vorticity $\overline{\omega}_z$ in the symmetry plane $z=0$ (first row) and of the streamwise vorticity $\overline{\omega}_y$ (second row) over a single period of bounce for $(Bo, Ga) = (0.05, 25)$. The corresponding evolutions for the case $(0.25, 30)$ (not shown) display similar features. During the bounce, vorticity is generated at the surface of the bubble and at the wall. In what follows, we refer to these two distinct contributions as surface and wall vorticity, respectively. Throughout the motion, the spanwise surface vorticity maintains the same sign from the front to the back of the bubble, indicating that no flow separation occurs at the bubble surface at such Bond numbers, even when the bubble gets very close to the wall. However, comparing panels $(b1)$ and ($d1)$ suggests that the surface vorticity is almost suppressed on the wall-facing side of the bubble when the gap becomes very thin, owing to the presence of a `tongue' of negative wall vorticity that almost cancels the positive surface vorticity. 
This tongue results from the structure of the flow present in the gap, in which the Bernoulli effect makes the vertical negative velocities (in the bubble's reference frame) larger near the bubble surface than near the wall, resulting in a negative velocity gradient, $\partial_x u_y < 0$, hence in a negative $\overline\omega_z$. Upstream and downstream of the bubble, the wall vorticity is positive, since the nearby fluid is entrained upwards (with respect to the wall). The comparison of panels $(c1)$ and $(g1)$, both of which correspond to the same gap thickness ($X_b \approx 1.25$) is of interest. The tongue of negative wall vorticity is much reduced in the latter, showing that, during the stage when the bubble is receding from the wall, the downward flow in the gap is inhibited by the vigorous upward entrainment of the near-wall fluid. Clearly, the significant positive wall-normal the bubble velocity is capable of sucking up the near-wall fluid located downstream, enabling it to catch up with the bubble.\\
\textcolor{black}{\indent Owing to the wall-induced asymmetry of the flow field, the bubble shape may exhibit a certain level of left-right asymmetry, particularly when it gets very close to the wall. 
When the minimum gap is very small ($\overline\delta=\mathcal{O}(0.01)$), such as in figures \ref{fig:regular_bounce_motion1} and \ref{fig:pretest1}, the overpressure resulting from lubrication effects flattens the part of the bubble surface located closest to the wall, as illustrated in figure \ref{fig:mesh}$(c)$. 
Conversely, when the minimum gap is one order of magnitude larger and the Bond number is `not too small', typically $Bo\gtrsim0.2$, the asymmetry changes sign. Specifically, at the smallest separation ($\overline\delta\approx0.125$) reached by the bubble with $(Bo, Ga)=(0.25, 30)$ (figure \ref{fig:regular_bounce_motion2}), the maximum curvature of the interface in the gap is $18\%$ larger than that on the fluid-facing side, indicating that the bubble is now more pointed on the wall-facing side. Since the flow in the gap is still dominated by inertial effects given the larger $\overline\delta$, the observed asymmetry may be ascribed to the Bernoulli effect, which decreases the pressure in the gap and thereby increases the curvature of the interface on the wall-facing side.}\\
\indent The second row in figure \ref{fig:regular_bounce_vor1} reveals the spatial distribution of the streamwise vorticity component in the half-space $z<0$. This component, which is antisymmetric with respect to the plane $z=0$, is concentrated within an elongated thread (hence, a second identical thread with opposite sign is present in the half-plane $z>0$). Pairs of elongated counter-rotating streamwise vortices are typical of the wake of axisymmetric bodies immersed in a shear flow, and were already observed in simulations in which a spherical bubble was steadily translating parallel to a nearby wall \citep{2020_Shi, 2024_Shi}. Here, the near-wall fluid is entrained upwards by the rising bubble, providing the ambient shear $\partial_x u_y\neq0$ required for the pair of streamwise vortices to be generated. Throughout the bubble ascent, $\overline\omega_y$ retains a constant sign in each half-plane, resulting in an entrainment of the fluid present in between the two $\overline\omega_y$-threads towards the wall. Following the classical shear-induced lift generation mechanism \citep{1956_Lighthill}, this entraiment yields a transverse force directed away from the wall. According to the succession of snapshots in figure \ref{fig:regular_bounce_vor1} , streamwise vorticity starts to appear at the rear of the bubble when the gap is close to its minimum (panel $(d2)$). Then, the two $\overline\omega_y$-threads grow continuously as the bubble departs from the wall (panels $(d2-h2)$), until the gap reaches its maximum (panel $(a2)$). As soon as the bubble starts to return towards the wall, the two threads are shed downstream with their tail bending towards the wall (panel $(b2)$), leaving the bubble wake free of streamwise vorticity (panel $(c2)$).\\
\indent The structure of the vorticity field revealed by figure \ref{fig:regular_bounce_vor1} is strongly time-dependent. However, since the gap varies from one panel to the other, one can wonder whether the observed variations are essentially enslaved to those of $\overline\delta(T)$, or if they are rather governed by the intrinsic unsteadiness inherent to the bounce dynamics. This question is examined in appendix \ref{sec:appD}, using separate computations carried out with a spherical bubble translating steadily at a given distance from the wall. Comparing the two sets of results establishes that the sequence described in figure \ref{fig:regular_bounce_vor1} has little to do with a quasi-steady evolution taking place at successive wall-bubble separations. This conclusion underlines the need to account for unsteady vortical effects in any model aimed at reproducing the main features of the bounce dynamics revealed by figure \ref{fig:regular_bounce_motion1}. In particular, the instantaneous force balance on the bubble must account for the existence of an inertial force involving the bubble acceleration, whose origin stands in the above time-dependent wake dynamics. We shall come back to this crucial point in \S\,\ref{sec:discuss}.

\subsection{Damped near-wall bouncing}
\label{sec:damp_bounce}

\begin{figure}
\vspace{5mm}
    \centering
    \includegraphics[scale=0.60]{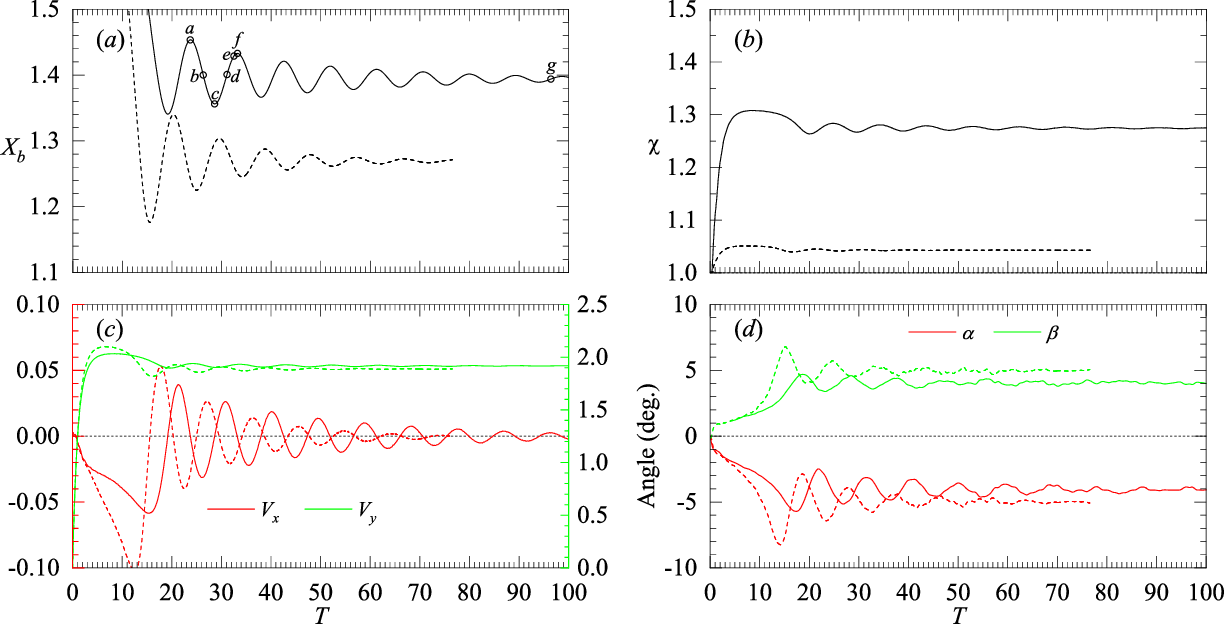}
    \caption{Evolution of various characteristics of the bubble and path during a damped bouncing sequence. Solid and dashed lines correspond to conditions $(Bo, Ga) = (0.25, 20)$ and $(0.05, 15)$, respectively. ($a$): wall-normal position (both bubbles are released from $X_0=2$, but the early evolution is not shown); ($b$): aspect ratio; ($c$): velocity of the bubble centroid, the left and right axes referring to $V_x$ and $V_y$, respectively; ($d$): inclination and drift angles.}
    \label{fig:damp_bounce_motion}
\end{figure}

Figure \ref{fig:damp_bounce_motion} displays the evolution of the bubble and path characteristics during the damped bouncing sequences observed for $(Bo, Ga) = (0.25, 20)$ (solid line) and $(0.05, 15)$ (dashed line). The gradual attenuation of the lateral path oscillations is evident in figure \ref{fig:damp_bounce_motion}$(a)$, as well as in figure \ref{fig:damp_bounce_motion}($c)$ which shows in particular the rapid damping of the wall-normal velocity. The decrease of $V_x$ is especially quick for the nearly-spherical bubble. Indeed, after a single quasi-period of bounce, the extreme values of $V_x$ are reduced by approximately $50\%$ in the case $(Bo, Ga) = (0.05, 15)$, and by $30\%$ for $(Bo, Ga) = (0.25, 20)$. Beyond this stage, the aspect ratio (panel $(b)$) and inclination (panel $(d)$) of the bubble quickly reach nearly constant values. In particular, the bubble is found to slightly incline towards the wall at an angle of approximately $5^{\circ}$ in both cases.

The final wall-normal position{\color{black}{, say $X_c$,}} at which the bubble stabilizes is governed by the relative magnitude of the attractive inviscid Bernoulli effect and the repulsive wake-wall interaction mechanism.  
Using the dimensionless terminal rise speed, $V_f$, the terminal Reynolds number may be expressed as $\Rey = 2Ga V_f$. In the low-$Bo$ case ($Bo=0.05$), figure \ref{fig:damp_bounce_motion}$(c)$ yields $\Rey=57$. At the same $\Rey$, the fixed-bubble simulations of \cite{2024_Shi} predict $X_c= 1.25$ for a spherical bubble, in good agreement with the final position $X_c\approx 1.27$ displayed in figure \ref{fig:damp_bounce_motion}$(a)$. In the second case ($Bo=0.25$), the equilibrium aspect ratio of the bubble is $1.275$ (figure \ref{fig:damp_bounce_motion}$(b)$) and its final Reynolds number is $77$. Owing to this significant deformation, the surface vorticity is larger than it would be at the same $\Rey$ if the bubble were spherical \citep{2007_Magnaudet}. Therefore, the repulsive component of the transverse force acting on the bubble is increased, yielding a larger equilibrium separation distance $X_c \approx 1.4$. Interestingly, the fixed-bubble simulation predicts $X_c= 1.17$ with a spherical bubble at the same Reynolds number. Therefore, it may be concluded that even a moderate deformation has a large effect on the transverse position at which the bubble stabilizes. \\
\begin{figure}
\vspace{5mm}
    \centering
    \includegraphics[scale=0.65]{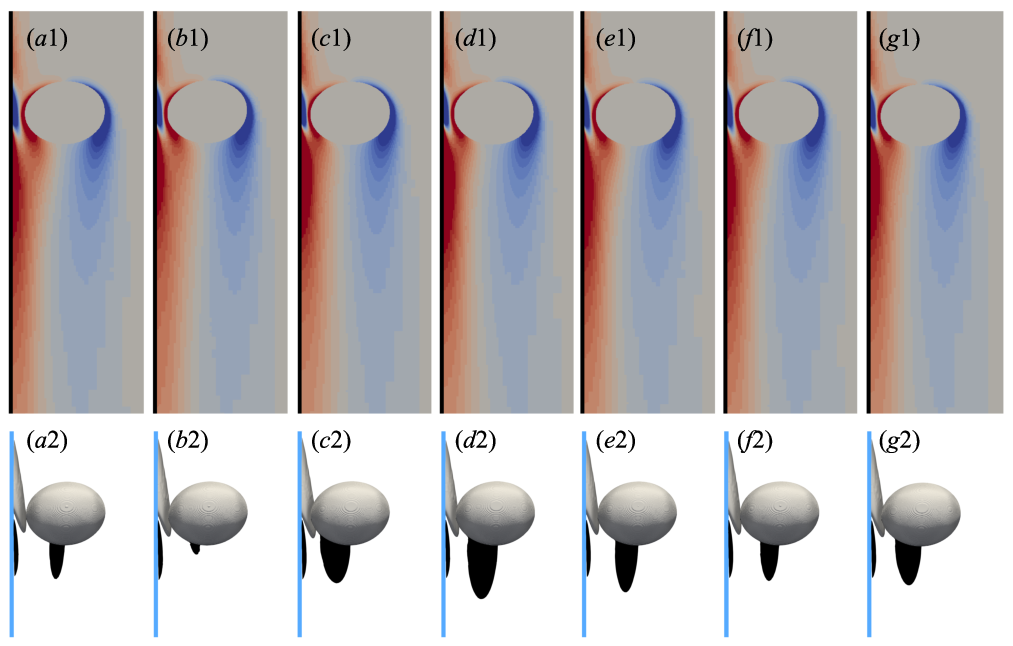}
    \caption{Same as figure \ref{fig:regular_bounce_vor1} for a bubble with $(Bo, Ga) = (0.25, 20)$. The respective instants of time are indicated by circles in figure \ref{fig:damp_bounce_motion}$(a)$. In the second row, the two iso-values correspond to $\overline\omega_y=\pm0.05$.}
    \label{fig:damped_bounce_vor1}
\end{figure}
\indent Figure \ref{fig:damped_bounce_vor1} illustrates the structure of the spanwise (first row) and streamwise (second row) component of the vorticity field past the bubble with $(Bo, Ga) = (0.25, 20)$ at the various instants of time specified in figure \ref{fig:damp_bounce_motion}$(a)$ (results for the case $(Bo, Ga) = (0.05, 15)$ display similar features). 
Throughout a single quasi-period of bouncing, the spanwise vorticity field (figure \ref{fig:damped_bounce_vor1}$(a1-f1)$) remains largely unchanged. Most notably, the region where the wall vorticity is positive appears to follow closely the evolution of the bubble position, rather than lagging significantly behind it as observed in figure \ref{fig:regular_bounce_vor1} during a periodic sequence of bounces. This difference stems from the contrasting evolutions of the rise speed in the two regimes: while $V_y(T)$ experiences little variations after the initial transient in the case of damped bounces (figure \ref{fig:damp_bounce_motion}$(c)$), it varies periodically by $15$ to $30\%$ in the periodic bouncing regime (figures \ref{fig:regular_bounce_motion1}$(c1)$ and \ref{fig:regular_bounce_motion2}$(c)$), which in turn results in significant `memory' effects on $\overline\omega_z(T)$ in this regime. 
In contrast to these modest changes, the streamwise vorticity field undergoes noticeable variations during a quasi-period of bouncing, as illustrated in the second row of figure \ref{fig:damped_bounce_vor1}. Its evolution shares some similarities with that observed in the periodic bouncing regime. In particular, at points $b$ and $d$ which are close to the equilibrium transverse position (almost reached at point $g$), $|\overline\omega_y|$ is weaker (resp. stronger) than in the equilibrium configuration when the bubble approaches (resp. departs from) the wall, as may be deduced by comparing figures \ref{fig:damped_bounce_vor1}$(b2)$ (resp. \ref{fig:damped_bounce_vor1}$(d2)$) and \ref{fig:damped_bounce_vor1}$(g2)$. 
However, the intensity of $\overline\omega_y$ is typically one order of magnitude weaker that in the periodic bouncing regime (compare the level of the iso-values in the bottom row of figures \ref{fig:regular_bounce_vor1} and \ref{fig:damped_bounce_vor1}). Furthermore, no evidence of a shedding process may be detected in the present case. 

\section{Additional insights into the bouncing regime}
\label{sec:discuss}
\subsection{Amplitude and frequency of the transverse oscillations}
\begin{figure}
\vspace{5mm}
    \centering
    \includegraphics[scale=0.65]{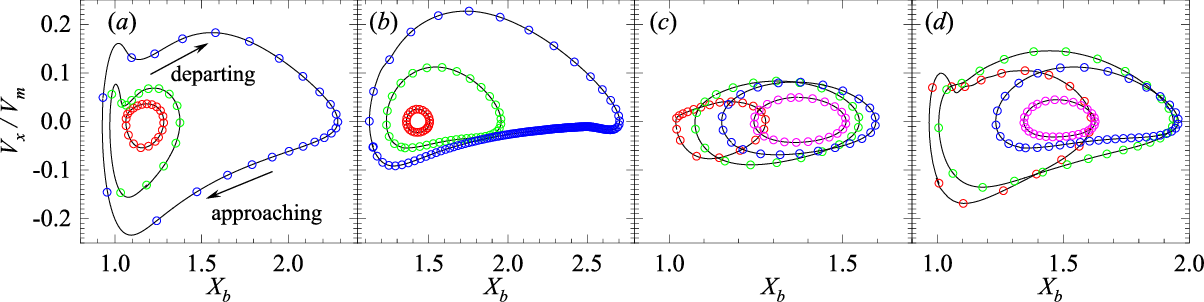}
    \caption{Variation of the bubble wall-normal velocity, $V_x$, normalised by the mean rise speed, $V_m$, as a function of the wall distance, $X_b$. In each series, the time interval between two adjacent points is $0.5$. ($a$) $\upchi \approx 1.03$ with $\Rey=73$ (\protect\textcolor{red}{$\circ$}), 118 (\protect\textcolor{green}{$\circ$}), 194 (\protect\textcolor{blue}{$\circ$}). ($b$) $\upchi \approx 1.4$ with $\Rey=100$ (\protect\textcolor{red}{$\circ$}), 123 (\protect\textcolor{green}{$\circ$}), 147 (\protect\textcolor{blue}{$\circ$}). ($c$) $\Rey \approx 80$ with $\upchi=1.02$ (\protect\textcolor{red}{$\circ$}), 1.12 (\protect\textcolor{green}{$\circ$}), 1.18 (\protect\textcolor{blue}{$\circ$}), 1.22 (\protect\textcolor{magenta}{$\circ$}). ($d$) $\Rey \approx 120$ with $\upchi=1.08$ (\protect\textcolor{red}{$\circ$}), 1.16 (\protect\textcolor{green}{$\circ$}), 1.42 (\protect\textcolor{blue}{$\circ$}), 1.49 (\protect\textcolor{magenta}{$\circ$}). Corresponding values of $Ga$ and $Bo$ are given in table \ref{tab:St}.}
    \label{fig:vx_2_vy}
\end{figure}

\indent Figure \ref{fig:vx_2_vy} shows the evolution of the normalised wall-normal bubble velocity against the transverse position of the bubble centroid over a single bouncing period. With nearly spherical or moderately deformed bubbles (panels $(a-b)$), the magnitude of the lateral oscillations increases continuously with the Reynolds number. For instance, with $\upchi \approx 1.03$ (panel $(a)$), their amplitude goes from $\approx0.11$ at $\Rey=73$ to $0.66$, i.e. six times more, at $\Rey=194$. 
These features remain qualitatively unchanged for $\upchi \approx 1.4$ (panel $(b)$), except that the equilibrium position is significantly further away from the wall. This avoids direct bubble-wall collisions, in contrast to what happens with $\upchi \approx 1.03$ at the highest two $\Rey$,  for which min$(X_b)<1$ according to panel $(a)$. 
That the equilibrium position shifts gradually towards the fluid interior as the bubble becomes more oblate is confirmed in panels $(c-d)$. This trend is consistent with the strengthening of vortical effects as $\upchi$ increases, and establishes the increase of the equilibrium separation as one of the primary consequences of increasing bubble deformation. An exception to this rule is the case $\upchi=1.49$ in panel $(d)$: as figure \ref{fig:chi-vs-re} indicates, this configuration is very close to the border with the non-oscillating regime in which the bubble migrates away from the wall. 
\begin{table}
\centering
\begin{tabular}{@{}l*{3}{l}c*{3}{l}c*{4}{l}c*{3}{l}@{}}
& \multicolumn{3}{c}{\raisebox{-1.5ex}{$\upchi\approx1.04$}} && \multicolumn{3}{c}{\raisebox{-1.5ex}{$\upchi\approx1.40$}} && \multicolumn{4}{c}{\raisebox{-1.5ex}{$\Rey\approx 80$}} && \multicolumn{3}{c}{\raisebox{-1.5ex}{$\Rey\approx 120$}} \\ 
\cmidrule(r){2-4} \cmidrule(lr){6-8} \cmidrule(lr){10-13} \cmidrule(l){15-17}
\Rey & 73 & 118 & 194 && 100 & {\color{black}{123}} & 147 && 83 & 84 & 82 & 79 && 121 & 120 & 127 \\
$\upchi$ & 1.02 & 1.03 & 1.04 && 1.40 & {\color{black}{1.42}} & 1.42 && 1.02 & 1.12 & 1.18 & 1.22 && 1.08 & 1.16 & 1.49 \\
$Ga$ & 18 & 25 & 30 && 25 & {\color{black}{28}} & 30 && 20 & 20 & 20 & 20 && 25 & 25 & 30 \\
$Bo$ & 0.02 & 0.02 & 0.015 && 0.3 & {\color{black}{0.25}} & 0.2 && 0.02 & 0.1 & 0.15 & 0.2 && 0.05 & 0.1 & 0.3 \\
$St$ & 0.127 & 0.116 & 0.066 && 0.105 & {\color{black}{0.062}} & 0.025 && 0.122 & 0.114 & 0.106 & 0.112 && 0.124 & 0.081 & 0.085 \\
$\overline{A}$ &0.22 &0.40 & 1.36&& 0.14&{\color{black}{0.73}}&1.58&& 0.26&0.48&0.45&0.27&& 0.64&0.94&0.29\\
$X_c$&1.17&1.17&1.60&&1.43&{\color{black}{1.60}}&1.91&&1.15&1.31&1.38&1.38&&1.29&1.47&1.49
\end{tabular}
\caption{Characteristics of the selected periodic bouncing configurations shown in figure \ref{fig:vx_2_vy}. The crest-to-crest oscillation amplitude, $A$, and bouncing frequency, $f$, are normalised in the form $\overline{A}=A/R$ and $St=2fR/V_m$, respectively.}
\label{tab:St}
\end{table}
\begin{figure}
    \centering
    \includegraphics[scale=0.65]{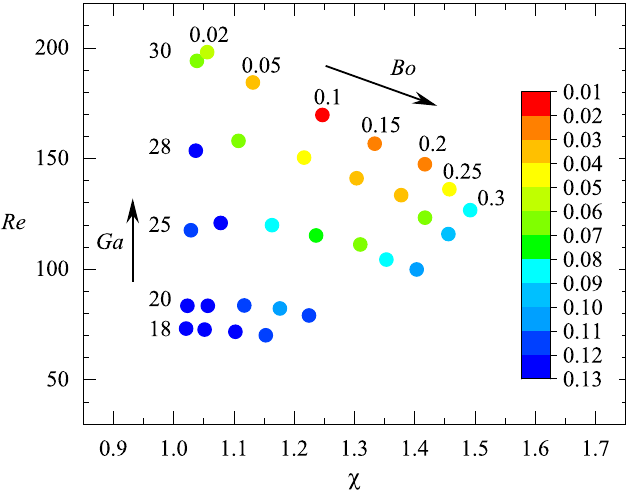}
    \caption{Reduced frequency $St$ of the oscillations throughout the periodic bouncing regime. The color map helps appreciate the variations of $St$ with $\Rey$ and $\upchi$.}
    \label{fig:st}
\end{figure}
Variations of the amplitude of the oscillations with $\upchi$ are more complex. For instance, in panel $(d)$ ($\Rey \approx 120$), this amplitude is seen to increase with the aspect ratio up to $\upchi=1.16$, whereas it decreases with $\upchi$ when the bubble is more oblate. According to figure \ref{fig:ini_sep_traj}, the initial transverse position at which the bubble is released does not influence significantly this amplitude. 
Therefore, one has to conclude that it is controlled by the nonlinearities of the local near-wall mechanisms that drive the bouncing motion. \\
\indent Figure \ref{fig:st} shows how the reduced bouncing frequency, $St=2fR/V_m$, varies with the control parameters. Values of $St$ for cases examined in figure \ref{fig:vx_2_vy} are given in table \ref{tab:St}. In the lowest two series of figure \ref{fig:st} ($Ga=18$ and $20$, or equivalently $Re\approx70$ and $80$), oscillations only occur with slightly deformed bubbles ($\upchi\lesssim1.2$, or $Bo\lesssim0.15$). Under such conditions, the reduced frequency does not vary much with the bubble oblateness ($St\approx0.115\pm0.01$). A different behaviour is observed for the upper three series, in which all configurations correspond to $Re>100$. Here, starting from a nearly spherical shape and increasing the Bond number in a given series up to $Bo\approx0.10-0.15$, the reduced frequency is found to decrease sharply, reaching values as low as $St\approx0.015$ for $(Bo,Ga)=( 0.1,30)$, i.e., $(\upchi,\Rey)\approx(1.25,170)$. Then, further increasing $Bo$, $St$ re-increases, recovering values of $\mathcal{O}(0.1)$ for $Bo=0.3$, i.e., $1.4\lesssim\upchi\lesssim1.5$, the maximum oblateness beyond which the periodic bouncing regime ceases. Influence of the Reynolds number on the oscillation frequency of nearly-spherical bubbles ($\upchi=1.03\pm0.01$) may also be appreciated by considering the left block of values in table  \ref{tab:St}. A continuous decrease of $St$ with $\Rey$ is noticed, from $St=0.126$ at $\Rey=73$ ($Ga=18$) to $St=0.067$ at $\Rey=194$ ($Ga=30$). \\
\begin{figure}
    \centering
    \includegraphics[scale=0.65]{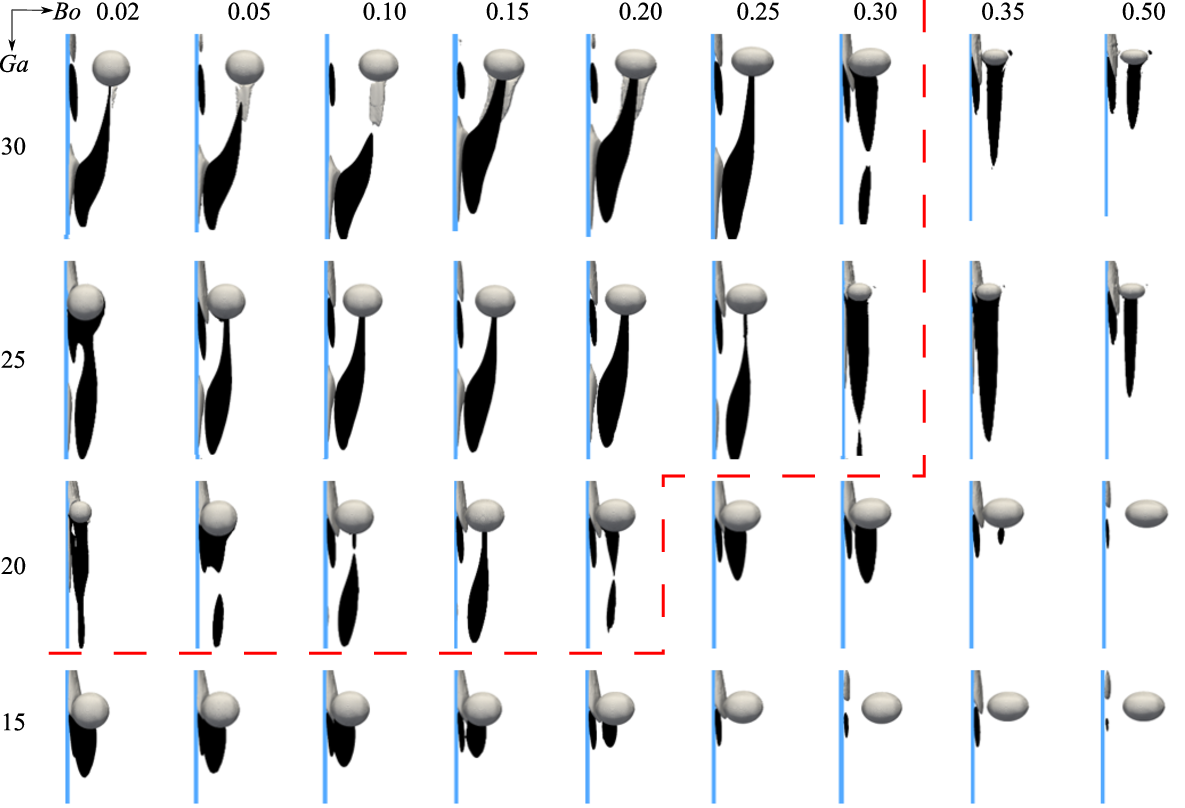}
    \caption{Iso-surfaces $\overline{\omega}_y= -0.5$ (black) and $+0.5$ (grey) of the streamwise vorticity in the half-space $z<0$ taken at the moment the streamwise vortices reach their maximum extension.  Maxima of $\overline{\omega}_y$ are much smaller in some cases, especially near the top right corner of the figure. In such cases, the threshold is reduced by a factor of $10$ and these snapshots provide a zoomed-out view in which bubbles appear smaller. The red dashed line marks the border between the periodic bouncing and damped bouncing regimes.}
    \label{fig:streamvort}
\end{figure}
\indent The above trend may be rationalized by returning to the mass-spring idealization of the system. In this framework, the net transverse force is considered to vary linearly with the transverse position $X_b$ in the vicinity of the equilibrium bubble position $X_c$, and the slope associated with this variation, $K_s(\Rey,\upchi)$, is the stiffness of the spring. The effective mass of the system that undergoes successive accelerations during bounces is proportional to the volume of fluid entrained by the bubble in its transverse motion. This  effective mass, say $M_v(\Rey,\upchi)$, varies over time, owing to the time variations of the wall-bubble separation and those of the bubble shape. Moreover, it also depends on the amount of fluid enclosed in the volume occupied by the streamwise vortices, as these time-dependent structures move with the bubble \citep{2006_Dabiri}. Since the mass-spring system oscillates with a dimensionless radian frequency $\Omega=2\pi St=(K_s/M_v)^{1/2}$, variations of the reduced frequency observed in figure \ref{fig:st} are due to those of $K_s$ and $M_v$ with $\Rey$ and $\upchi$. The experimental data of TM (their figure 7) suggest that $K_s$ varies only slowly with the Reynolds number in the case of nearly-spherical bubbles. In contrast, one expects the strength of the streamwise vortices to grow significantly with $\Rey$ at a given $\upchi$ (or with $Ga$ at a given $Bo$). To check this point, snapshots of the $\overline\omega_y$-distribution throughout the parameter range $Ga\geq15$, $Bo\leq0.5$ are plotted in figure \ref {fig:streamvort}. This figure fully confirms the above guess, the size of the streamwise vortices at the moment they reach their maximum extension exhibiting a marked and gradual growth with $Ga$ in each series. This growth translates directly into an increase of the effective mass $M_v$, yielding a decrease of $St$ with $Ga$ (or $\Rey$) if $K_s$-variations remain small, in line with the numerical findings reported in figure \ref{fig:st}. To the best of our knowledge, no detailed determination of the net transverse force has been achieved so far with significantly deformed bubbles, so that variations of $K_s$ with $\Rey$ for aspect ratios larger than approximately $1.1$ are unknown. Nevertheless, the gradual decrease of $St$ with $Ga$ is observed whatever $Bo$ in figure \ref{fig:st}, making us believe that the above argument remains valid, even for $Bo\gtrsim0.1$. The lack of knowledge regarding the variations of $K_s$ with $\upchi$ at a given Reynolds number also makes it difficult to rationalise the variations of $St$ with $\upchi$ (or $Bo$). However, a partial clue may be extracted from the above results. First, figure \ref{fig:streamvort} does not reveal any significant variation in the size of the streamwise vortices with the Bond number, at least up to $Bo\approx0.15$. Therefore, one has to conclude that $M_v$ does not change much with the bubble oblateness for such modest deformations. In contrast, as discussed above, figure \ref{fig:vx_2_vy} and table \ref{tab:St} establish that, at a given $\Rey$, the oblateness controls the equilibrium position $X_c$, shifting it towards the fluid interior as $\upchi$ increases. Since the vortical and irrotational contributions to the transverse force vary less steeply with the bubble position as the separation increases (keep in mind that both decrease approximately as $X_b^{-n}$, with $n<2\leq4$ according to TM), the stiffness $K_s$, which is the slope of the net transverse force in the vicinity of $X_c$, is expected to decrease as $\upchi$ increases. Combining the above arguments suggests that $St\equiv(2\pi)^{-1}(K_s/M_v)^{1/2}$ is a decreasing function of $\upchi$ for low-to-moderate deformations, which figure \ref{fig:st} confirms whatever $Ga$. The above arguments are not sufficient to explain the re-increase of $St$ with $Bo$ observed for larger oblatenesses in the upper three series ($Ga\geq25$). This surprising trend results presumably from the combined variations of $K_s$ and $M_v$. Only a specific study of the near-wall variations of the transverse force as a function of $\upchi$ for a fixed oblate spheroidal bubble may help rationalize completely the complex variations of $St$ with the bubble shape in the future.
\subsection{Connection between regular near-wall bouncing and wall vorticity}
\label{sec:key_feature}

Section \ref{sec:regular_bounce} established the existence of a vortex shedding process in the periodic bouncing scenario observed for the specific parameter sets $(Bo,Ga) = (0.05,25)$ and $(0.25,30)$. To get more insight into the connection between vortex shedding and periodic bouncing, we examined the evolution of $\overline\omega_y(T)$ for all $(Bo,Ga)$ sets in which the bubble motion follows this scenario, i.e., all configurations located above and on the left of the dashed red line in figure \ref {fig:streamvort}. This examination confirmed that the shedding process takes place in all cases. Conversely, no vortex shedding is observed in the other two regimes. Since the pair of streamwise vortices directly contributes to the repulsive transverse force on the bubble, it appears that the regular bouncing motion is closely related to the flow characteristics that make this shedding possible. According to figure \ref{fig:traj_sum}$(a)$, periodic bouncing is only observed with Bond numbers less than $0.3$, hence with nearly spherical or moderately deformed bubbles. In an unbounded domain where the fluid is at rest, the wake of such bubbles is stable \citep{2007_Magnaudet}, and therefore remains axisymmetric. Hence, the generation of the streamwise vortex pair in the presence of a wall is made possible by the velocity gradients present in the gap, which result from the no-slip condition. More specifically, given the finite span of the bubble, the vertical fluid velocity is nonuniform in the spanwise direction $(\partial_zu_y\neq0)$. Therefore, the spanwise wall vorticity $\omega_z^w \approx\partial_xu_y\big|_{x=0}$ associated with the near-wall shear flow induces a vortex tilting term, $\omega_z^w\partial_z u_y$, resulting in a nonzero streamwise vorticity component, $\omega_y$. 
This generation mechanism prompts to an examination of the connection between $\omega_z^w$ and the shedding of the vortex pair.
\begin{figure}
\vspace{5mm}
\centering
\includegraphics[scale=0.65]{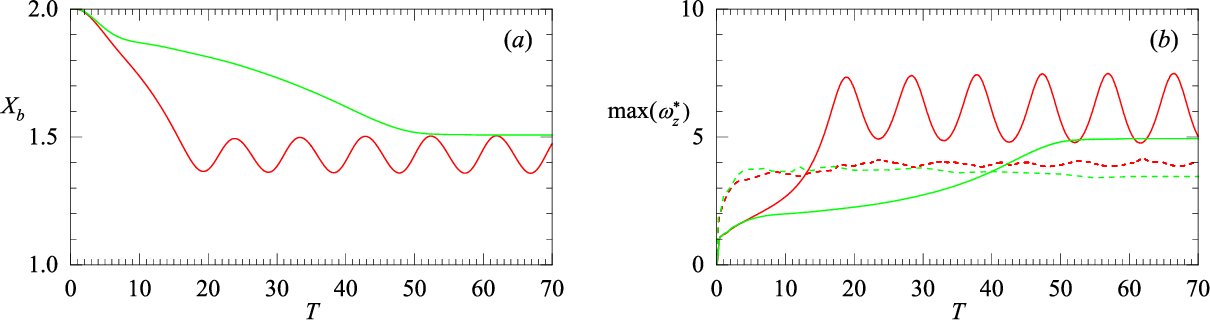}
\caption{Evolution of some characteristics of the bubble and fluid motions for the parameter sets $(Bo,Ga)=(0.3, 25)$ (red lines) and $(0.35, 25)$ (green lines). $(a)$: normalised wall-normal bubble position $X_b(T)$; ($b$): peak values of the normalised surface vorticity $|\omega^{*}{_z^s}|(T)$ (dashed lines), and wall vorticity $|\omega^{*}{_z^w}|(T)$ (solid lines). }
\label{fig:vor-history}
\end{figure}

\begin{figure}
\vspace{5mm}
\centering
\includegraphics[scale=0.75]{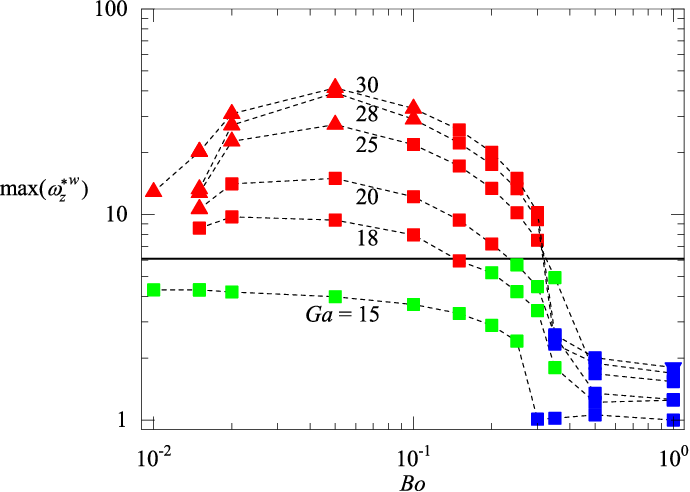}
\caption{Variation of the maximum  normalised wall vorticity, $|{\omega^{*}_{z}}^{w}|$, with the Bond and Galilei numbers. Symbols are identical to those of figure \ref{fig:traj_sum}, with \textcolor{red}{$\blacktriangle$} and \textcolor{red}{$\blacksquare$} referring to the periodic bouncing regime (with and without wall-bubble collisions, respectively), \textcolor{blue}{$\blacktriangledown$} and \textcolor{blue}{$\blacksquare$} referring to lateral migration away from the wall (with and without path instability, respectively), and \textcolor{green}{$\blacksquare$} corresponding to the damped bouncing regime. Values of $Ga$ increase from bottom to top. In cases involving direct wall-bubble collisions, $|{\omega^{*}_{z}}^{w}|$ is estimated in the approaching stage at the moment when $\overline\delta\approx5\overline\Delta_\text{min}$. The thick horizontal line corresponds to max$(|{\omega^{*}_{z}}^{w}|)= 6.0$.}
\label{fig:max-vorw}
\end{figure}
To gain some insight into this relationship, we examine the evolution of the peak values of the wall vorticity $\omega_z^w$ and the surface vorticity $\omega_z^s$ for the two parameter sets $(Bo, Ga) = (0.3,25)$ and $(0.35, 25)$. Although the two sets have aspect ratios and Reynolds numbers with close values ($(\upchi, Re)= (1.40, 99.8)$ and $(1.48, 99.0)$, respectively), the first case belongs to the periodic bouncing regime while the second corresponds to an over-damped bouncing motion, as illustrated in figure \ref{fig:vor-history}$(a)$. Figure \ref{fig:vor-history}$(b)$ compares the evolutions of the normalised wall and surface spanwise vorticity components, both of which are made non-dimensional here by the advective time scale $R/||\boldsymbol{v}(t)||$, implying $\omega^{*}{_z^w}= 2\overline\omega_z^wGa/Re$ and $\omega^{*}{_z^s}= 2\overline\omega_z^sGa/Re$. Since the vertical fluid velocity is close to $||\boldsymbol{v}(t)||$ just ahead of and behind the bubble, i.e., at a distance from the wall of the order of $RX_b(T)$,  $|\omega^{*}{_z^w}|(T)$ may be thought of as a measure of the inverse of the instantaneous non-dimensional bubble-wall separation. While $\omega^{*}{_z^s}(T)$ (dashed lines) remains relatively constant and keeps similar values in both cases, the two evolutions of $\omega^{*}{_z^w}(T)$ reveal striking differences. Specifically, the magnitude of $\omega^{*}{_z^w}$ stabilizes around a value of $5.0$ for $(Bo,Re)=(0.35,25)$. In contrast, it undergoes regular oscillations for $(Bo,Ga)=(0.3,25)$, peaking at roughly $7.5$ when the bubble is closest to the wall, a value almost twice as large as that of $\omega^{*}{_z^s}$. These observations suggest that the shedding of streamwise vortices occurs only in the presence of large enough  $\omega^{*}{_z^w}$.
This suggestion is further validated in figure \ref{fig:max-vorw} which gathers the maximum normalised wall vorticity, $\max(|\omega^{*}{_z^w}|)$, for all cases studied in this work. It is observed that in all periodic bouncing cases,  the maximum of $|\omega^{*}{_z^w}|$ exceeds a value of $6.0$. Keeping $Ga$ fixed, this maximum is seen to peak at $Bo=0.05$, irrespective of $Ga$. Therefore, the shedding of streamwise vortices in the bubble wake, itself a key ingredient of the periodic bouncing regime, is confirmed to be closely associated with the existence of a sufficiently intense  $\omega^{*}{_z^w}$, i.e., of a sufficiently small minimum wall-bubble separation. 

\section{Summary and concluding remarks}
\label{sec:summary}

In this work, we conducted time-dependent three-dimensional resolved numerical simulations to study the rising motion of a single deformable bubble released near a vertical hydrophilic wall in a quiescent liquid. Our investigation essentially focused on the moderately inertial regimes in which bubbles ascend in a straight line when the fluid domain is unbounded. 
In line with available experimental findings, this investigation confirms that three distinct types of near-wall paths may take place in the range of parameters we considered, namely migration away from the wall, damped near-wall oscillations, and periodic near-wall bouncing. More specifically, the physical picture that emerges from the simulations is as follows:
\begin{itemize}
    \item If the Galilei number is lower than a critical $Bo$-dependent value, $Ga_1(Bo)=\mathcal{O}(10)$ (or equivalently a Reynolds number less than $\Rey_1\approx35$), or if the Bond number is larger than a weakly-$Ga$-dependent critical value, $Bo_2(Ga)\approx0.35-0.4$, the scene is dominated by the vortical mechanism associated with the tiny flow correction induced by the interaction of the wake with the wall at large distances downstream of the bubble. Under such conditions (which correspond to blue symbols in figures \ref {fig:traj_sum}$(a)$ and \ref{fig:chi-vs-re}), the bubble is consistently repelled from the wall throughout its ascent and whatever the initial distance separating it from the wall.
\item If $Ga>Ga_1$ and $Bo<Bo_2$, the transverse force acting on the bubble is dominated down to a short $(Ga,Bo)$-dependent distance to the wall, $X_c$, by the attractive Bernoulli mechanism predicted by potential flow theory and resulting from the acceleration of the flow in the gap. Below this critical separation, the near-wall vortical activity is strong enough to keep the net transverse force locally repulsive. Nevertheless, the consequences of the near-wall repulsive mechanisms may take two different forms:\\
\indent - If the critical separation $X_c(Bo,Ga)$ is large enough because viscous effects are still significant in the bulk, the bubble reaches this position in a quasi-steady manner. Under such conditions, it may enter the near-wall repulsive region, but does not get close enough to the wall for vorticity in the gap to be sufficiently intense to trigger vortex shedding. Hence, the wake remains quasi-stationary (although fully three-dimensional) and the transverse motion damps gradually. This is the picture encountered in the range $\Rey_1\leq\Rey\leq\Rey_2\approx65$ with nearly-spherical bubbles. Since wake effects are reinforced by the bubble deformation, this regime subsists up to higher Reynolds numbers for bubbles having intermediate Bond numbers $Bo_1(Ga)\leq Bo\leq Bo_2(Ga)$, with $Bo_1$ increasing in the range $0.2-0.35$ with $Ga$ (green symbols in figures \ref {fig:traj_sum}$(a)$ and \ref{fig:chi-vs-re}).\\
\indent - On the contrary, if viscous effects in the bulk are weak enough for the bubble to keep a sufficient transverse velocity when it approaches the critical separation, it may get so close to the wall that the normalised wall vorticity $\omega^{*}{_z^w}$ defined in \S\,\ref{sec:key_feature} exceeds the threshold value of $6.0$. Then, vortex shedding sets in, providing a repulsive contribution capable of driving the bubble back to its previous extreme lateral position in the attractive region, despite the viscous drag that opposes this departing motion. This is the essence of the mechanisms that make the existence of a periodic bouncing regime possible. This regime is encountered for $Ga\geq Ga_1\approx18$ and $Bo\leq Bo_1(Ga)$, which corresponds to bubbles with a low-to-moderate deformation, the maximum oblateness varying from $1.15$ at $Ga=18$ to $1.5$ at $Ga=30$ (red symbols in figures \ref {fig:traj_sum}$(a)$ and \ref{fig:chi-vs-re}).
\end{itemize}
 \indent It is important to note that, in contrast to bouncing phenomena near a horizontal or slightly inclined wall, time variations of the bubble shape do not appear to play a central role when the wall is vertical. These variations are significant, bubbles becoming less (resp. more) oblate when they get very close to (resp. depart from) the wall, as the results discussed in \S\,\ref{sec:regular_bounce} establish. However, we showed that this evolution is partly a consequence of the variations of the bubble rise speed, which is forced to decrease when the gap becomes very thin and to re-increase when it widens, owing to the no-slip condition at the wall. Of course, these changes in the bubble shape alter all quantities influenced by the body geometry, such as the effective mass of fluid it entrains laterally, the viscous resistance involved in the transverse motion, or the attractive contribution to the transverse force. Nevertheless, their consequences on the transverse motion appear to be minor compared to those of the vortex shedding process.\\
\indent As additional simulations performed with fixed spherical bubbles confirmed, this vortex shedding phenomenon is specific to freely-moving bubbles having a transverse motion with respect to the wall. Indeed, in the parameter range considered here, the flow past weakly or moderately deformed bubbles held at a fixed distance from the wall is strictly stationary. When the bubble is maintained close enough to the wall, the no-slip condition forces the flow in the gap to be strongly sheared. The classical vortex tilting mechanism then leads to a wake dominated by the presence of a pair of counter-rotating streamwise vortices in which fluid particles rotate in such a way that the fluid located in between the two vortices is deflected towards the wall, yielding a repulsive transverse force. 
With bouncing bubbles, the same generation mechanism holds and the two streamwise vortices keep the same orientation as in the fixed-bubble configuration, but they undergo a periodic shedding cycle intimately related to the transverse motion of the bubble. More specifically, the vortices start to form when the wall-bubble gap reaches its minimum, and grow continuously  throughout the departing stage of the bounce. Shedding takes place when the gap reaches its maximum, and the wake then remains virtually free of streamwise vorticity throughout the approaching stage. Hence, the shedding cycle is not locked to the instantaneous transverse velocity of the bubble (which vanishes at both extremities of the lateral oscillations), nor to its acceleration (which vanishes near the centreline of the oscillations). Rather, its dynamics appears to depend in a complex manner on the past history of the transverse motion, making the repulsive force provided by the streamwise vortices look like a `memory' effect.\\
\indent Results of the present study help clarify several of the physical mechanisms responsible for the different regimes of the wall-induced migration of a deformable bubble. 
These results may be used to develop predictive low-order models capable of reproducing the observed path characteristics in all three regimes. Clearly, most ingredients of such models are already qualitatively known, although their detailed variations with respect to the Reynolds number have most of the time been determined only for spherical or nearly-spherical bubbles. This is the case of the virtual mass force predicted by potential flow theory and of the viscous drag force associated with the transverse motion, which may both be reasonably estimated as a function of the wall-bubble separation. Similarly, the quasi-steady inertial  transverse force has been determined for spherical bubbles held at fixed distances from the wall \citep{2020_Shi,2024_Shi}. Therefore, what is essentially lacking at present is a model capable of properly modulating this force as a function of time, so as to mimic its actual time variations imposed by the vortex shedding process. The bubble kinematics provided by this study over a significant range of flow conditions may be used to build heuristically such a model. For this, knowing the bubble position, velocity and acceleration at every instant of time, and assuming known approximate expressions for all other forces, the time-dependent vortex-induced force could be estimated, and its evolution could be connected to those of the transverse velocity and acceleration to derive empirical `kernel' functions.

\vspace{2mm}

\noindent{\bf Funding} {P.S. acknowledges the funding of the Deutsche Forschungsgemeinschaft (DFG, German Research Foundation) through grant number 501298479. }
\vspace{2mm}

\noindent{\bf Declaration of interests} {The authors report no conflict of interest.}
\vspace{2mm}

\noindent{\bf Author ORCIDs} \\
{
\noindent Pengyu Shi, https://orcid.org/0000-0001-2345-6789\\
Jie Zhang, https://orcid.org/0000-0002-2412-3617\\
Jacques Magnaudet, https://orcid.org/0000-0002-6166-4877.
}


\appendix
\section{Numerical tests for $(Bo,Ga) = (0.073,21.9)$}
\label{sec:appA}

In a first series of preliminary tests, we focused on the parameter set $(Bo,Ga) = (0.073,21.9)$, aiming at replicating the near-wall bouncing motion observed by TM.

\begin{figure}
	\centerline{\includegraphics[scale=0.65]{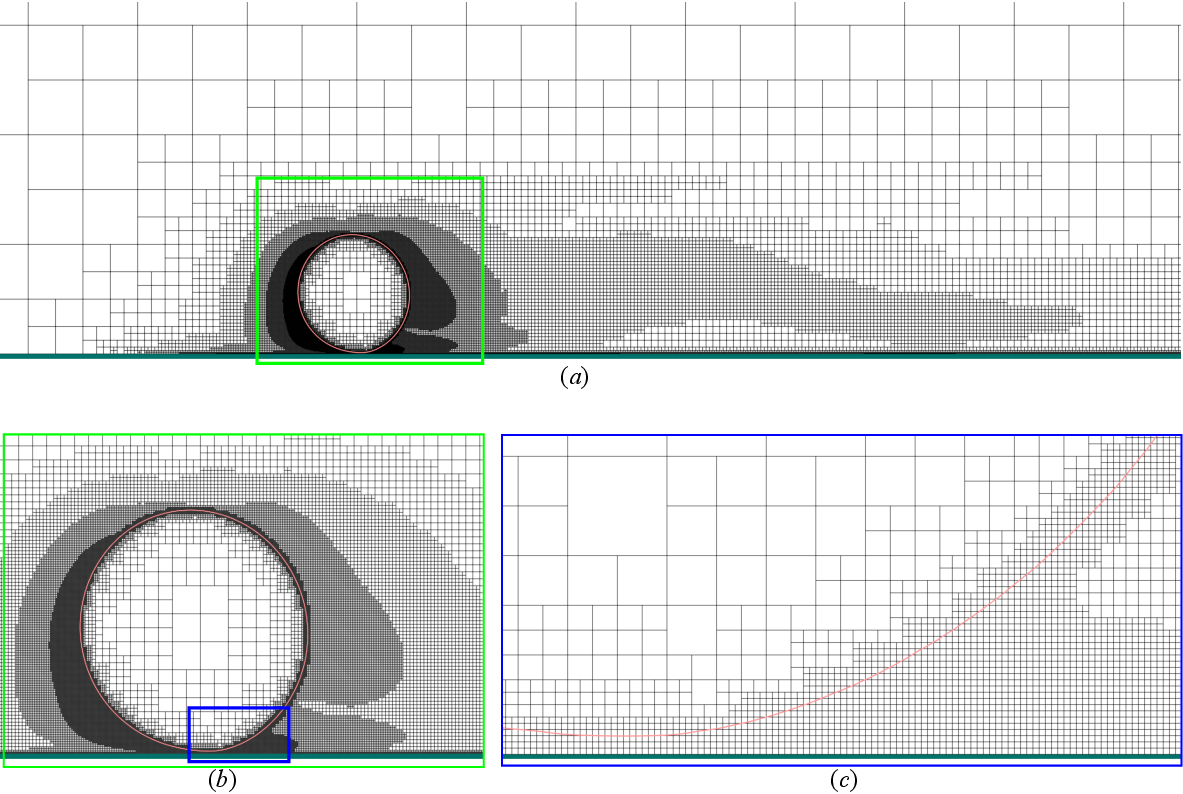}}
\caption{Illustration of the grid structure in the symmetry plane $z=0$ slightly before the gap reaches its minimum ($\Delta_\text{min}=1/136$). In each panel, the bubble rises from right to left. Its surface is marked with a red line; the wall is indicated by a dark green line at the bottom.}
\label{fig:mesh}
\end{figure}

The accuracy of the simulations carried out with \emph{Basilisk} significantly depends on the CFL number, $N_{CFL}$, the standard tolerance $T_\epsilon$ on the Poisson solver providing the pressure field, the grid refinement criteria for the volume fraction $\zeta_C$ and the velocity $\zeta_u$, and the dimensionless minimum grid size $\overline\Delta_{\min}$. In all considered cases, we set $\zeta_C = 10^{-3}$, which ensures that all grid cells near the bubble surface are refined using $\overline\Delta_{\min}$. The reference case was run with $N_{CFL} = 0.5$, $T_\varepsilon = 10^{-4}$, $\zeta_u = 1\times10^{-2}$, and $\overline\Delta_{\min} = 1/68$. \citet{2021_Zhang, 2022_Zhang}, who considered the problem of two identical bubbles initially rising inline for $Ga$ up to 90 and $0.02 \leq Bo \leq 1$, showed that these numerical settings, together with $\zeta_C = 10^{-3}$, provide reliable predictions. An example of the corresponding grid structure in the symmetry plane $z=0$ for the present problem is illustrated in figure~\ref{fig:mesh}. The grid is captured at the moment when the bubble is closest to the wall. Due to the very narrow gap, the minimum grid size is reduced to $\overline\Delta_{\min} = 1/136$ instead of $\overline\Delta_{\min} = 1/68$ (the reason behind this choice will be discussed later) to improve the local resolution.\\
      \begin{figure}
		\centerline{\includegraphics[scale=0.65]{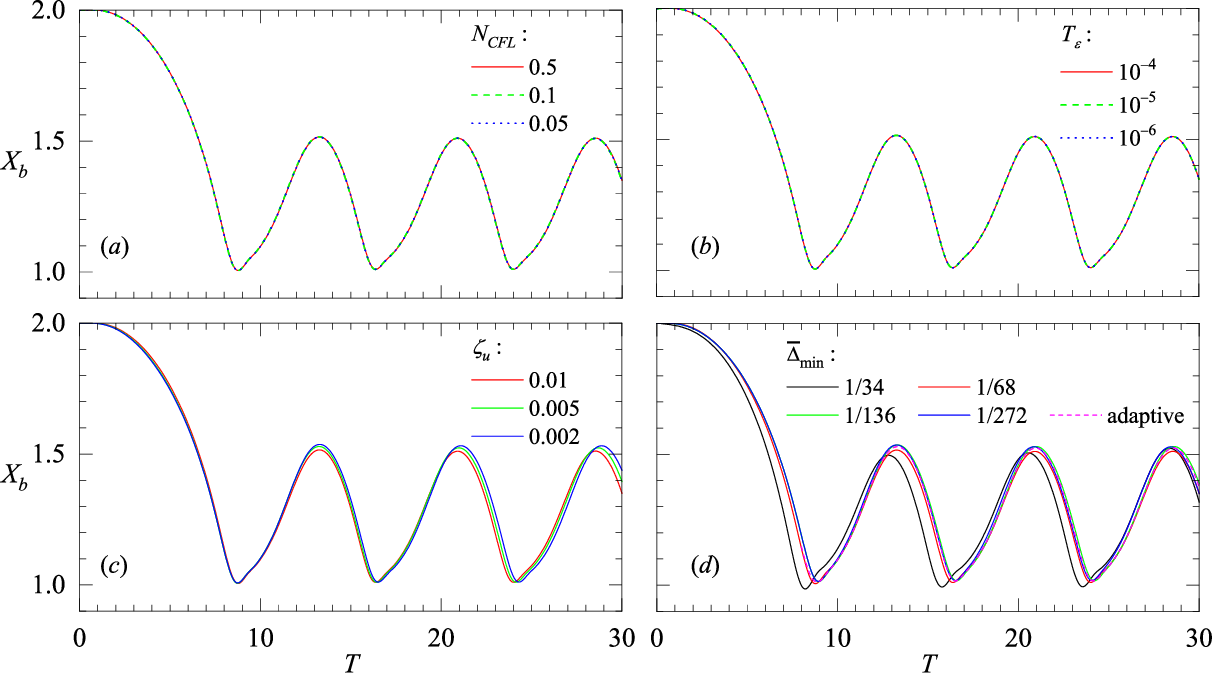}}
		\caption{Effects of ($a$): $N_{CFL}$; ($b$): $T_\varepsilon$; ($c$): $\zeta_u$; and ($d$): $\Delta_{\min}$, on the predicted lateral position $X_b(T)$. \textcolor{black}{($e$): effect of $\Delta_{\min}$ on the gap thickness $\overline\delta(T)$; the color code is similar to that of panel $(c)$ and the successive $\Delta_{\min}$ are materialized by the horizontal dashed lines.} The bubble centroid is initially located at $X_0 = 2$. }
		\label{fig:pretest1}
	\end{figure}
\indent Experimental results (TM) indicated that the corresponding motion is symmetric with respect to $z = 0$, a conclusion confirmed by the results we obtained in the base test case mentioned above. This motivated us to consider only a half domain in subsequent simulations and to use a symmetry boundary condition on the plane $z = 0$. Figures \ref{fig:pretest1}$(a-d)$ summarize the results obtained for the time evolution of the wall-normal bubble position $X_b(T)$. While the prediction is hardly affected by any further decrease in either $N_{CFL}$ or $T_\varepsilon$, significant variations take place with decreasing $\zeta_u$ (figure \ref{fig:pretest1}$(c)$) and $\overline\Delta_{\min}$ (figure \ref{fig:pretest1}$(d)$). These variations are no surprise: using a smaller $\zeta_u$ increases the resolution in the bubble wake, while a smaller $\overline\Delta_{\min}$ improves the resolution in the gap, a crucial region of the flow when the bubble gets very close to the wall. For the considered parameter set, the Reynolds number is around 100, so that the dimensionless thickness of the bubble boundary layer is about $0.1$. With $\overline\Delta_{\min}= 1/68$ and $\zeta_u=10^{-2}$, more than six grid points are usually located in the immediate vicinity of the bubble, ensuring a sufficient resolution of the boundary layer. Decreasing $\zeta_u$ from $10^{-2}$ to $2\times10^{-3}$, the typical non-dimensional size of the grid cells within the far wake (located about ten radii downstream of the bubble) is reduced from $1/17$ to $1/34$. This refinement only leads to a slightly larger maximum of $X_b$ after each bounce. However, achieving this minor improvement is computationally expensive, as the total number of grid cells increases from 1.1 million to 7 million, even when considering only a half domain. This is why, to save computational ressources, we decided to use $\zeta_u=10^{-2}$ in all runs. {\color{black}{Note that the small staircases that may be discerned in the iso-contours in the wake region in figures \ref{fig:depart_vor2}$(g1-i1)$ and similar are a direct consequence of this compromise in the choice for $\zeta_u$, and no longer exist with $\zeta_u=2\times10^{-3}$.}}\\
\indent More pronounced variations in the evolution of $X_b(T)$ are observed when changing $\overline\Delta_{\min}$, highlighting the importance of achieving a sufficient spatial resolution in the gap. As indicated in figure \ref{fig:pretest1}$(d)$, while the bouncing frequency is weakly affected, the predicted amplitude of the lateral displacement of the bubble centroid during each bounce exhibits a sizeable increase when reducing $\overline\Delta_{\min}$ from $1/68$ to $1/136$\textcolor{black}{, resulting in a $5\%$ increase of the crest-to-crest amplitude of the oscillations.} The good agreement between the results obtained with $\overline\Delta_{\min}=1/136$ and those with $\overline\Delta_{\min}=1/272$ suggests that using $\overline\Delta_{\min}=1/136$ is sufficient to achieve an almost grid-independent solution \textcolor{black}{on macroscopic quantities, such as $X_b(T)$}. \textcolor{black}{Nevertheless, lubrication effects at stake when the bubble moves very close to the wall start to be reasonably captured only with $\overline\Delta_{\min}=1/272$.} \textcolor{black}{Indeed, as figure \ref{fig:pretest1}$(e)$ reveals, the minimum gap thickness increases dramatically when $\overline\Delta_\text{min}$ is decreased from $=1/68$ to $1/136$, eventually stabilizing at $\overline\delta\approx0.013$ with $\overline\Delta_{\min}=1/272$. The very small minimum, $\overline\delta\approx2\times10^{-3}$, observed with the coarsest resolution is well below  the corresponding $\overline\Delta_{\min}$, leading to a situation of `direct collision'. This situation is avoided with $\overline\Delta_{\min}=1/136$, but only one and a half cells lie in the film when the gap reaches its minimum. Last, three grid cells are located in the gap with $\overline\Delta_{\min}=1/272$, which is a bare minimum to resolve the semi-Poiseuille flow therein.} \\
\indent To further reduce computational costs, we adjust $\overline\Delta_{\min}$ depending on the separation distance, so that the grid is more refined only when the bubble moves very close to the wall. Therefore, $\overline\Delta_{\min}$ is decreased from $1/68$ to $1/136$ when the dimensionless gap $\overline\delta$ is less than $0.15$. This ensures that the number of cells in the gap is always larger than ten for $\overline\delta\geq0.075$. The comparison in figure \ref{fig:pretest1}$(d)$ indicates that this adaptive refinement strategy (the results of which are labelled as \emph{adaptive} in the figure) is able to faithfully replicate the results obtained by {\color{black}{prescribing $\overline\Delta_{\min}=1/136$ from the beginning of the simulation.}}\\
\indent The above discussion indicates that the numerical error may be minimized at a reasonable computational cost by setting $N_{CFL} = 0.5$, $T_\varepsilon = 10^{-4}$, $\zeta_u = 1\times10^{-2}$, and lowering $\overline\Delta_{\min}$ from $1/68$ to $1/136$ only when $\overline\delta(T)$ is less than $0.15$. The reliability of this parameter set may be assessed by comparing the corresponding predictions with the experimental results of TM. Figure \ref{fig:pretest2}($a$) shows the predicted Reynolds number $Re$ based on the bubble rise speed as a function of time. The corresponding mean value, averaged over a single bouncing period, is approximately 96.09, closely aligning with the measured value of 96.25. Figure \ref{fig:pretest2}($b$) presents a similar comparison for the bubble aspect ratio $\upchi$. According to the prediction, $\upchi$ decreases down to $1.08$ when the bubble is closest to the wall and increases to $1.14$ when it achieves its maximum separation. These extreme values align well with the measured ones.
	 \begin{figure}
		\centerline{\includegraphics[scale=0.65]{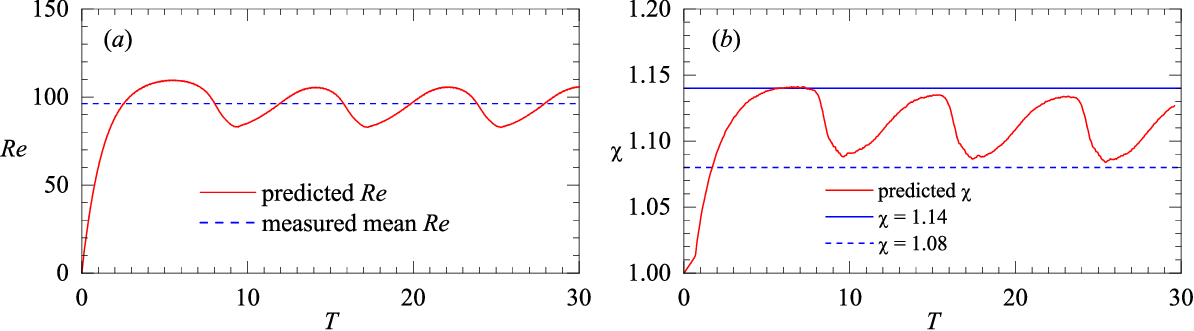}}
		\caption{Results for ($a$): the Reynolds number $Re$; and ($b$): aspect ratio $\upchi$. In ($b$), the two blue lines highlight the maximum and minimum $\upchi$ observed in the experiment.}
		\label{fig:pretest2}
	\end{figure}	
    In the above experiments, the evolution of the wall-normal distance $X_b(T)$ was found to be accurately fitted as (in present notations)
	\begin{equation}
	X_b(T) =X_c+\overline{A} \sin\Omega T\,, \quad\mbox{with\   }
	X_c=1.3, \quad \overline{A} =0.256,\quad \Omega =0.379 
	\end{equation}
In contrast, in the numerical prediction reported in figure \ref{fig:pretest1}, the parameters of the oscillatory bubble motion are
\begin{equation}
	X_c=1.29, \quad \overline{A} =0.27,\quad \Omega =0.781 
\end{equation}
Compared with the experimental determinations, the mean transverse position $X_c$ and the amplitude $\overline{A} $ are replicated satisfactorily. In contrast, the predicted bouncing frequency appears to be twice as large as the measured value. To ascertain the source of this overestimate, we investigated the effects of certain parameters that may significantly influence the bouncing frequency. We first noted that the initial dimensionless separation $X_0$ in the experiment is around $1.4$, which is smaller than that in our tests. To determine whether this difference might contribute to the discrepancy, we performed additional runs at $X_0=1.5$ and $1.25$. As observed in figure \ref{fig:pretest3}($a$), although the duration required for the bubble to approach the wall decreases significantly with decreasing $X_0$, the subsequent oscillatory motion is barely affected by $X_0$. 	
Another potential cause of discrepancy could be a slight tilt of the vertical wall with respect to the vertical, leading to a decrease in the attractive transverse force, owing to the nonzero projection of the buoyancy force in the wall-normal direction. To check this effect, we increased the  tilt angle up to $-\theta = 2^\circ$ (figure \ref{fig:pretest3}($b$)), with $\theta$ defined to be positive if the wall-normal projection of the buoyancy force points towards the wall. While this change reduces the bouncing frequency, thereby reducing the disagreement with the observed $\Omega$, it significantly increases the amplitude of the oscillations, yielding an obvious overestimate on $\overline{A} $.\\
\begin{figure}
\vspace{5mm}
\centerline{\includegraphics[scale=0.65]{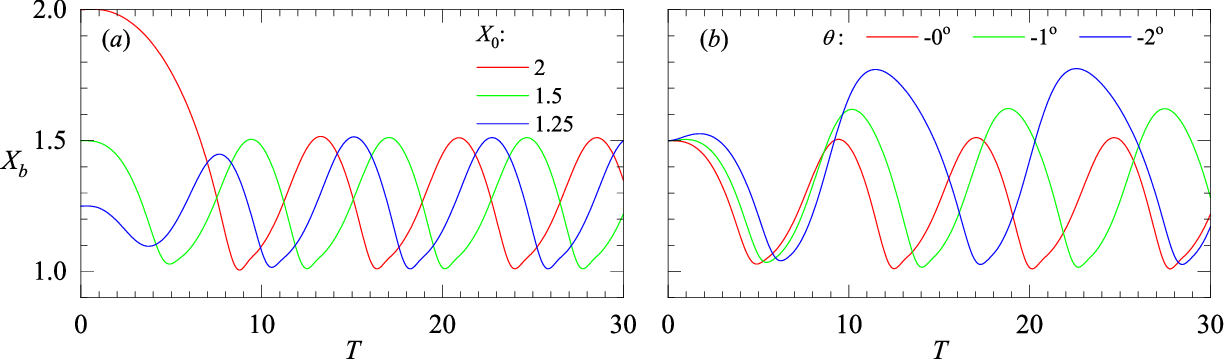}}
\caption{Influence of two parameters on the bouncing frequency. ($a$): effect of the initial separation $X_0$; $(b$): effect of a small wall inclination with respect to the vertical. }
\label{fig:pretest3}
\end{figure}
\indent Continuing with geometrical flow parameters, we questioned the possible influence of the boundary conditions at the wall on the bouncing frequency. Up to this point, the wall was assumed to be smooth and to impose a no-slip condition to the fluid. We explored two variants of this boundary condition. One maintains the no-slip condition but assumes the wall to be rough, while the other maintains the wall perfectly flat but assumes the fluid to obey a free-slip condition on it. In the first variant, the rough wall is conceptualized as a two-dimensional sinusoidal wave in the $(x,y)$ plane, described as $X = 0.1 |\sin5\pi Y|$. Figure \ref{fig:pretest4}($a$) provides a sketch of the corresponding configuration. Figure \ref{fig:pretest4}($b$) compares the resulting lateral motion of the bubble with that predicted in the reference case. It may be seen that the wall roughness, as described in this simple approach, has only a minimal influence on the bouncing frequency. In contrast, switching from the no-slip condition to the free-slip one makes the near-wall bubble motion transition from the regular bouncing regime to a regime in which the bubble quickly reaches a stable equilibrium position a significant distance apart from the wall.

\begin{figure}
\centerline{\includegraphics[scale=0.65]{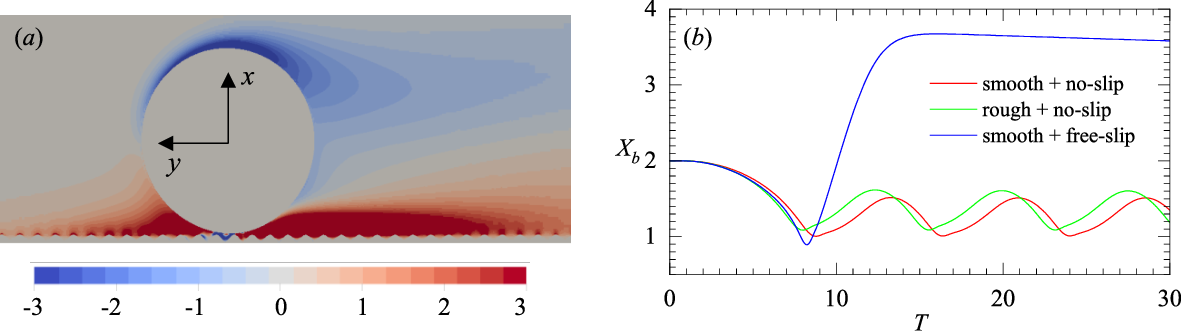}}
\caption{Influence of the wall roughness and possible fluid slip at the wall on the bubble lateral motion. ($a$): `rough' wall, modelled as a sinusoidal wave in the $(x,y)$ plane (with the spanwise vorticity $\overline{\omega}_z$ shown with coloured iso-contours); $(b)$: prediction of the lateral bubble motion resulting from the different boundary conditions at the wall. }
\label{fig:pretest4}
\end{figure}

\begin{figure}
\vspace{5mm}
\centerline{\includegraphics[scale=0.65]{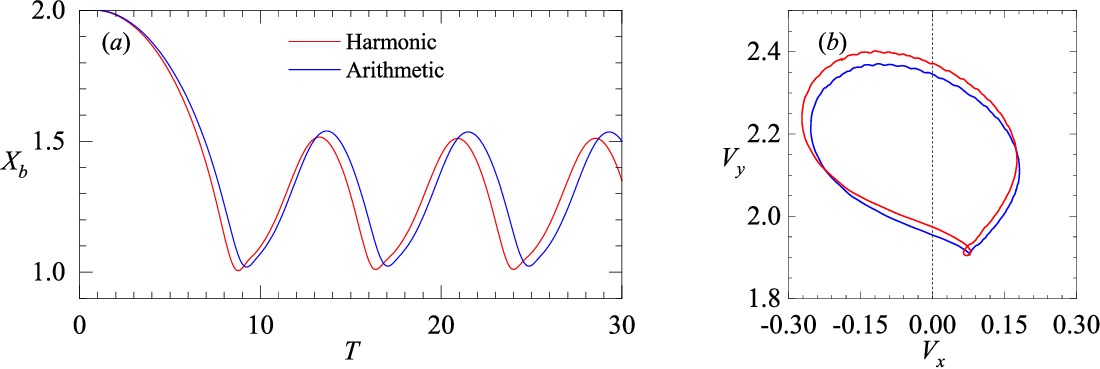}}
\caption{Influence of the interpolation rule (harmonic mean vs arithmetic mean) used to estimate the dynamic viscosity of the two-phase mixture. ($a$): time history of the bubble wall-normal position, $X_b$; ($b$): velocity diagram built on the vertical ($V_y$) and wall-normal ($V_x$) components of the bubble velocity over a full period of the bouncing motion.}
\label{fig:pre-test-mu}
\end{figure}

Having found no satisfactory explanation to the discrepancy on the bouncing frequency by changing several physical and geometrical parameters of the system, we also questioned some numerical aspects. In particular, we examined the possible influence of the empirical interpolation rule used to evaluate the dynamic viscosity of the fluid near the bubble surface. In all tests presented above, the harmonic weighting described by (\ref{eq:vof}$b$) was used. Alternatively, the arithmetic weighting, similar to that used for density in (\ref{eq:vof}a), is routinely employed in many studies. As noted by \citet{2011_Tryggvason}, the primary distinction between the two approaches is that the arithmetic mean tends to 'favour' the largest viscosity, yielding for instance $\mu(C=1/2)\approx\mu_2/2$ if the viscosity of fluid 2 is much larger than that of fluid 1, whereas the harmonic mean 'favours' the smallest viscosity, yielding $\mu(C=1/2)\approx2\mu_1$ in the same case. The influence of the choice of the interpolation rule for $\mu(C)$ on the lateral motion of the bubble is underscored in figure \ref{fig:pre-test-mu}. According to figure \ref{fig:pre-test-mu}$(a)$, employing the arithmetic mean slightly increases the duration of the initial transient required for the bubble to come close to the wall, and results in a marginally larger wall-bubble mean separation during the bouncing cycle. Figure \ref{fig:pre-test-mu}$(b)$ presents the velocity diagram of the bubble over a full cycle. It shows that the maximum vertical velocity achieved with the arithmetic weighting is somewhat lower than that resulting from the harmonic weighting (by nearly $1.5\%$), while the maximum negative horizontal velocity predicted with the second option is $6\%$ larger than that obtained with the harmonic weighting. These observations align with the intuitive idea that the arithmetic weighting induces slightly more pronounced viscous effects. However, differences between the two predictions remain small, and certainly not significant enough to cause a significant overestimate in the bouncing frequency.\\
\indent In summary, none of the possibilities considered above is responsible for the discrepancy on the bouncing frequency noticed with respect to the experimental results of TM. Turning to experiments, and excluding a typo since figure 9 of TM provides a record of $X_b$ vs the dimensional time $t$, it is unlikely that the discrepancy may be due to some contamination of the bubble surface by impurities, since silicone oils are non-polar. Similarly, the influence of a possible non-Newtonian behaviour of the oil in the narrow gap separating the bubble from the wall, where the shear rate is large, is unlikely. Indeed, silicone oils subjected to high shear rates are shear-thinning, which would tend to lower viscous effects, hence to promote the bouncing motion. Non-hydrodynamic interactions between the bubble surface and the wall, which add a `disjoining' contribution to the pressure field in the gap, seem more realistic candidates since the minimum gap is only a few microns. Of course, testing this possibility is beyond the scope of the present work. Instead, we decided to conduct additional tests on the motion of a single bubble close to an inclined or horizontal wall, to make sure that our purely hydrodynamic numerical predictions are reliable. The results of these tests and the comparison with available experimental data are detailed in appendix \ref{sec:appB}.

\section{Tests with a bubble bouncing near an inclined or horizontal wall}
\label{sec:appB}

We ran several additional tests in situations where an air bubble bounces near a flat horizontal or inclined wall. We first considered the case of a bubble rising up to a horizontal, hydrophilic wall, selecting the parameter set $(Bo, Ga) = (0.074, 63)$. This corresponds to an air bubble with an equivalent radius $R = 0.74\, \text{mm}$ rising in hyper-clean water, as considered experimentally by \citet{2014_Kosior}. Experimental results indicate that the flow field exhibits an axial symmetry. Therefore, we ran the simulations by constraining the flow to remain axisymmetric with respect to the minor axis of the bubble. This adaptation allowed us to refine the grid down to $\overline\Delta_{\min}=1/544$, making it possible to further examine the grid independence of the predictions.

	 \begin{figure}
	 \vspace{5mm}
		\centerline{\includegraphics[scale=0.65]{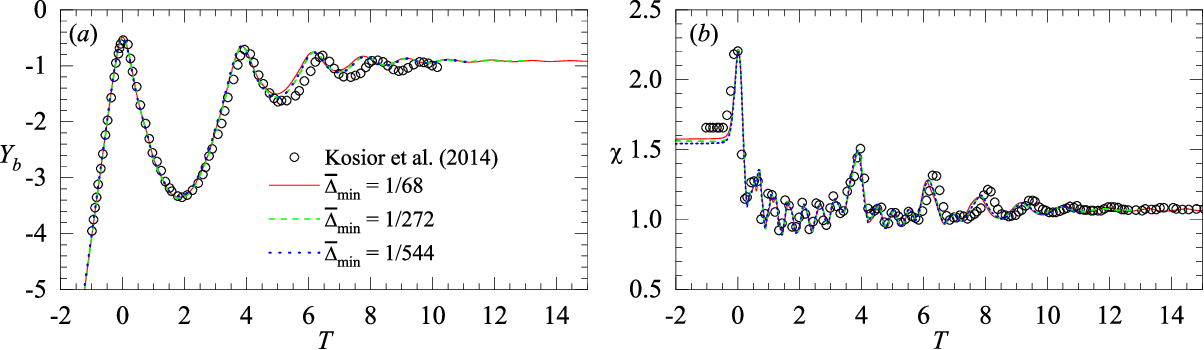}}
		\caption{Evolution of a bubble with $(Bo, Ga) = (0.074, 63)$ rising towards a horizontal wall. ($a$): wall-normal position of the bubble centroid; ($b$): geometrical aspect ratio. In both panels, $T=0$ corresponds to the moment when the wall-bubble distance reaches its first minimum.}
		\label{fig:pre-test-horizon_wall}
	\end{figure}
	
Figure \ref{fig:pre-test-horizon_wall} depicts the evolution of the normalised wall-normal position of the bubble centroid, $Y_b$, and the bubble aspect ratio, $\upchi$.  The predicted $\upchi$ (figure \ref{fig:pre-test-horizon_wall}$(b)$) is seen to closely match the experimental results, suggesting that a grid refinement down to only $\overline\Delta_{\min}=1/68 $ is sufficient to capture the dynamics of the bubble deformation. Figure \ref{fig:pre-test-horizon_wall}($a$) reveals a slight underestimate of the maximum separation the bubble achieves after the second collision, especially with $\overline\Delta_{\min}=1/68$. Nevertheless, this small difference does not significantly affect the predicted bouncing frequency. More precisely, the dimensionless time duration between the third and fourth collisions is approximately $1.78$ in the experiment. In the simulations, this time duration is approximately $1.65$ for $\overline\Delta_{\min}=1/68 $ and $1.675$ for $\overline\Delta_{\min}=1/544 $. In both cases, the relative difference with the experiments is less than $8\%$, much smaller than that observed in the presence of a vertical wall.

\begin{figure}
\centerline{\includegraphics[scale=0.65]{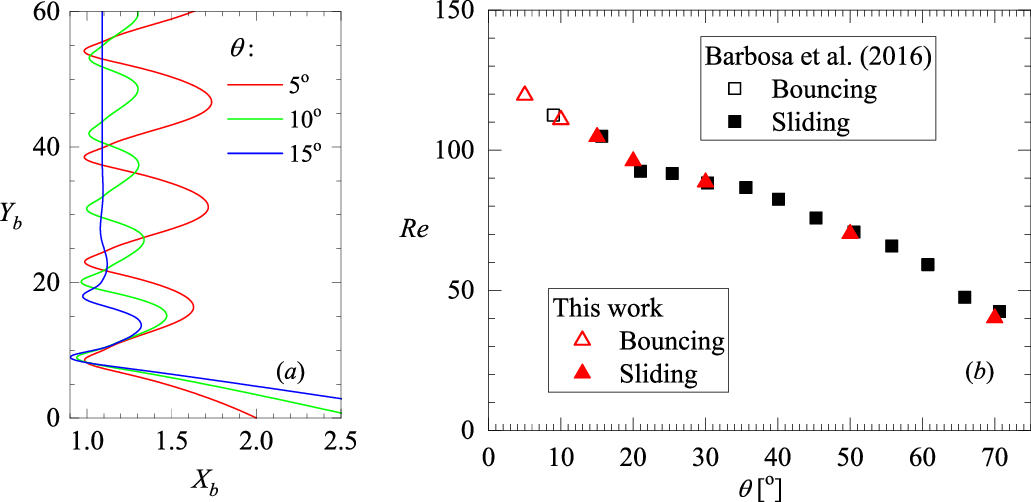}}
\caption{Evolution of a bubble with $(Bo, Ga) = (0.14, 27)$ rising along an inclined wall. ($a$): bubble trajectory; ($b$) Reynolds number based on the time-averaged bubble velocity in the wall-parallel direction. In ($a$), $X_b$ and $Y_b$ denote the normalised wall-normal and wall-parallel positions of the bubble centroid, respectively.}
\label{fig:pre-test-inclined_wall}
\end{figure}

The second configuration we considered corresponds to a clean bubble rising close to an inclined wall, as experimentally investigated by \citet{2016_Barbosa}. We focus on the parameter set $(Bo, Ga) = (0.14, 27)$, corresponding to an air bubble with an equivalent radius $R = 0.55\,\text{mm}$ rising in a silicone oil slightly more viscous than water but having a surface tension four times less than that of water (case E1 in \citet{2016_Barbosa}). The wall inclination angle, $\theta$, defined to be positive if the wall-normal projection of the buoyancy force points towards the wall, was varied from $5^\circ$ to $70^\circ$ ($\theta = 0^\circ$ corresponds to a vertical wall). \citet{2016_Barbosa} showed that the bubble trajectory transitions from a near-wall bouncing regime to a sliding regime beyond a critical inclination angle $\theta_{c}$, with $10^\circ<\theta_{c}<15^\circ$. Present predictions perfectly agree with these findings. In particular, in figure \ref{fig:pre-test-inclined_wall}($a$) the bubble is seen to bounce regularly for $\theta=5^\circ$ and $10^\circ$, while the oscillations of its centroid are quickly damped and the bubble later slides along the wall for $\theta=15^\circ$, maintaining a constant gap with $\overline\delta\approx0.1$. Figure \ref{fig:pre-test-inclined_wall}($b$) shows how the Reynolds number based on the time-averaged bubble velocity along the wall-parallel direction varies with the wall inclination. Again, numerical predictions are in excellent agreement with experimental data. 

\section{Influence of initial separation}
\label{sec:appC}
\begin{figure}
	\centerline{\includegraphics[scale=0.65]{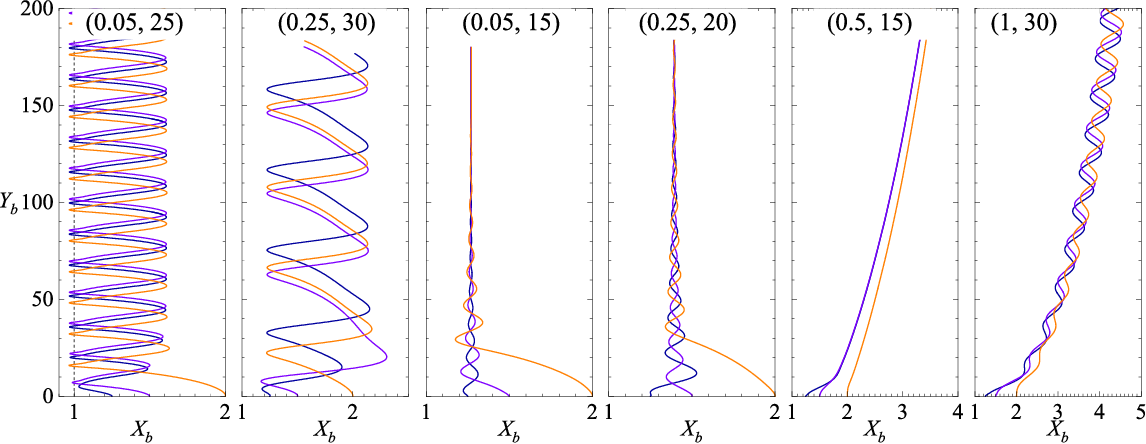}}
	\caption{Bubble trajectories corresponding to different initial separations $X_0 = 2$ (orange), $1.5$ (purple), and $1.25$ (dark blue). The corresponding $(Bo, Ga)$ are indicated at the top of each panel.}
	\label{fig:ini_sep_traj}
\end{figure}

\begin{figure}
\vspace{5mm}
	\centerline{\includegraphics[scale=0.65]{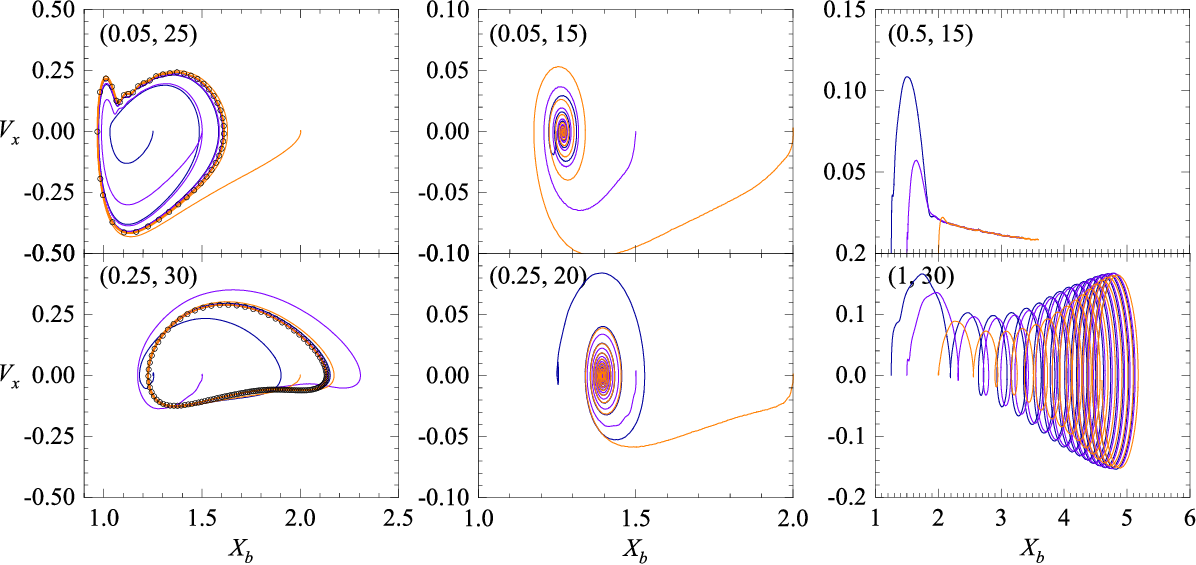}}
	\caption{Same as figure \ref{fig:ini_sep_traj} for the wall-normal velocity $V_x$ of the bubble centroid. In the left two panels, open circles correspond to the evolution of $V_x(X_b)$ in the fully developed stage; the time duration between two successive symbols is $0.2$.}
	\label{fig:ini_sep_ux}
\end{figure}

\begin{figure}
	\centerline{\includegraphics[scale=0.65]{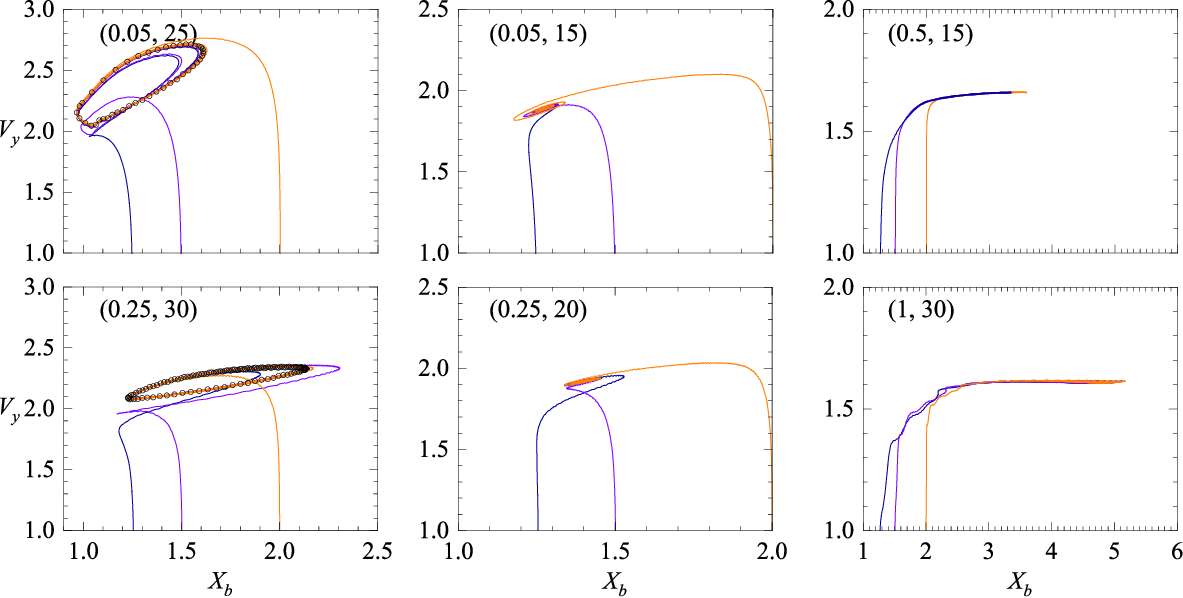}}
	\caption{Same as figure \ref{fig:ini_sep_traj} for the vertical velocity $V_y$ of the bubble centroid. In each panel, the portion where $V_y<1$ is omitted, as $V_y$ increases rapidly in the initial stage and therefore depends only weakly on $X_b$.}
	\label{fig:ini_sep_uy}
\end{figure}

The results discussed in the main body of the paper were obtained with an initial bubble-wall separation $X_0 = 2$. However, the conclusions drawn from these results are valid irrespective of the initial separation, provided that it is `not too large' for wall effects to remain sizeable. To support this claim, we examined the six typical cases discussed in \S\,\ref{sec:path_wake}, comparing the results obtained with initial separations $X_0 = 2$, $1.5$, and $1.25$. Figures \ref{fig:ini_sep_traj}, \ref{fig:ini_sep_ux}, and \ref{fig:ini_sep_uy} show the bubble trajectories and the evolution of $V_x(X_b)$ and $V_y(X_b)$ in these three cases, respectively. \\
In the parameter range where the bubble performs periodic bounces ($(Bo,Ga)=(0.05, 25)$ and $(0.25,30)$), the predicted bubble trajectories (figure \ref{fig:ini_sep_traj}) differ only within the first period of bouncing. In particular, the maximum wall-normal position that the bubble achieves during this early stage depends on $X_0$. In contrast, the bubble motion reaches a fully developed state from the second period of bouncing, and the vertical displacement and amplitude of the lateral drift during a single period then remain unaffected by $X_0$. Accordingly, in this stage, the profiles of the wal-normal and vertical velocities of the bubble centroid, $V_x(X_b)$ and $V_y(X_b)$, collapse on a single curve (denoted with open circles in figures \ref{fig:ini_sep_ux} and \ref{fig:ini_sep_uy}, respectively). This establishes the rapid memory loss of bubbles with respect to $X_0$ in the bouncing regime.

In the two damped bouncing configurations, $(Bo, Ga)=(0.05, 15)$ and $(0.25,20)$, the initial separation affects the vertical displacement achieved by the bubble before it stabilizes at the rest position $X_c$. Close inspection indicates that the smaller the initial offset $|X_0-X_c|$, the shorter the vertical displacement needed. This is confirmed in figure \ref{fig:ini_sep_ux} where, in each quasi-period of the motion, the curve with the smallest initial offset is seen to stay closest to the fixed point $(X_b=X_c, V_x =0)$ where all curves eventually concentrate. The same feature may be observed in figure \ref{fig:ini_sep_uy}, the fixed point then being $(X_b=X_c, V_y = V_f)$, with $V_f$ the final rise speed.

Last, in the two cases where the bubble departs from the wall, the effects of the initial separation depend on the specific details of the bubble motion. In the case where no path instability takes place ($(Bo,Ga)=(0.5, 15)$), the maximum $V_x$ achieved in the initial stage increases with decreasing $X_0$ (figure \ref{fig:ini_sep_ux}). 
Beyond the initial stage, 
the wall-normal and vertical velocity components are seen to depend solely on the local wall-normal distance, $X_b$. Therefore, although the trajectories do not overlap, the path of a bubble with a given $X_0$ is just a replication of the path of another bubble released at a larger $X_0$. 
In the marginal case $(Bo, Ga)=(1,30)$, a bubble rising in an unbounded expanse of fluid exhibits a zigzagging path \citep{2016_Cano-Lozano,2023_Bonnefis_b}. In the presence of a vertical wall, the mean lateral drift is driven by the wake-wall vortical interaction mechanism. Conversely, the oscillations in the lateral drift are essentially due to the path instability. This may be appreciated in figure \ref{fig:ini_sep_ux} where, beyond $X_b \approx 3$, the maximum and minimum of the wall-normal velocity achieved in each period of bouncing have almost the same magnitude. These results also indicate that the larger $X_0$, the smaller the wall-induced asymmetry in the flow, hence the longer it takes for the path instability to saturate (consider the growth of the oscillations of the red curve corresponding to $X_0=2$ in figure \ref{fig:ini_sep_ux} ). Of course, once the saturated state is reached, the characteristics of the motion are no longer affected by $X_0$, as figures \ref{fig:ini_sep_ux} and \ref{fig:ini_sep_uy} evidence.

 \section{Wake structure past a spherical bubble translating steadily close to a wall}

\label{sec:appD}

\begin{figure}
\vspace{5mm}
    \centering
    \includegraphics[scale=0.65]{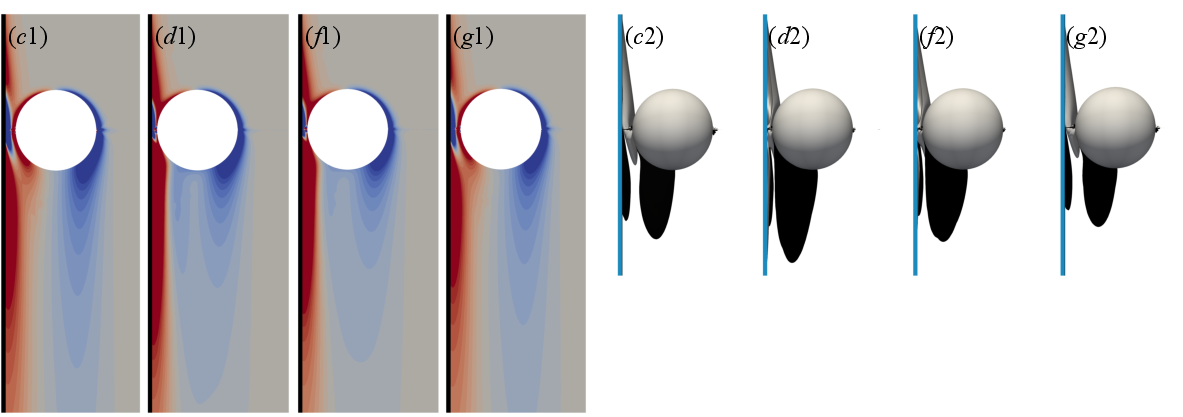}
    \caption{Vortical structure past a spherical bubble moving steadily parallel to a rigid wall in a stagnant fluid. $(\Rey, X_b)$ =(131, 1.25) in panels $(c1)-(c2)$, (126, 1.125) in $(d1)-(d2)$, (106, 1.125) in $(g1)-(g2)$, and (112, 1.25) in $(f1)-(f2)$. These $(\Rey, X_b)$ sets are similar to those of the freely-moving bubble at points $c, d, g$, and $f$ in figure \ref{fig:regular_bounce_motion1}$(a2)$. The color bar used for the $\overline\omega_z$ iso-contours (left four panels) and the selected iso-values for $\overline\omega_y$ (right four panels) are consistent with those of figure \ref{fig:regular_bounce_vor1}.}
    \label{fig:fixed_bubble_vor}
\end{figure}

To gain some additional insight into the dynamic evolution of the vortical structures during a regular bouncing, we ran simulations about a spherical bubble held fixed in a wall-bounded uniform flow. The parameters $\Rey$ and $X_b$ were selected to match those corresponding to points $c, d, f$, and $g$ in figure \ref{fig:regular_bounce_motion1}$(a2)$. These simulations were carried out using the in-house JADIM code developed at IMFT. The corresponding numerical details may be found in \citet{2020_Shi} and \citet{2024_Shi}. \\
\indent Figure \ref{fig:fixed_bubble_vor} presents the structure of the vorticity field obtained in this stationary configuration. The distribution of the spanwise vorticity (first four panels), closely mirrors that obtained with a freely moving bubble (see the panels with the corresponding labels in figure \ref{fig:regular_bounce_vor1}), except that the extension of the wall region with significant positive vorticity is larger in the fixed-bubble configuration. 
In contrast, the structure of the streamwise vorticity field (last four panels of figure \ref{fig:fixed_bubble_vor}), differs notably from that observed past a freely-moving bubble. Specifically, the intensity of $\overline\omega_y$ at points $c$ and $d$  is markedly higher in the case of a fixed bubble. This difference arises because the wake requires a certain time to `respond' to the vortex stretching/tilting process resulting from the shear flow present in the gap. This time lag is especially evident at point $c$, where $\overline\omega_y$ appears minimal for a freely moving bubble (figure \ref{fig:regular_bounce_vor1}$(c2)$), while $\overline\omega_z$ exhibits significant values in the gap at the same instant of time (figure \ref{fig:regular_bounce_vor1}$(c1)$). Conversely, the streamwise vorticity at points $f$ and $g$ is more pronounced with a freely-moving bubble, althoug the shear flow in the gap is weaker than in the fixed-bubble case (compare panels $(f1)-(g1)$ in the two figures). 

The above differences in the two $\overline\omega_y$-distributions highlight the importance of unsteadiness in the instantaneous vorticity distribution past the bubble. This influence translates into significant `memory' effects (i.e., effects that depend on the past history of the transverse motion) in the lateral force balance on the bubble. Since the streamwise vortices displayed in figure~\ref{fig:fixed_bubble_vor}$(c2-g2)$ correspond to the classical Lighthill lift mechanism, they result in a stationary force directed away from the wall. For a freely-moving bubble experiencing bounces, the time-dependent generation/disappearance of the streamwise vortices yields an attractive (resp. repulsive) `memory' correction to this repulsive stationary force during the time the bubble approaches (resp. departs from) the wall. Consequently, at the equilibrium transverse position where the quasi-steady transverse force vanishes, its transient counterpart is attractive in the approach stage and repulsive in the return stage, prompting the bubble to always deviate from its equilibrium position. 


\bibliographystyle{jfm}
\bibliography{bubble}

\begin{thebibliography}{51}
\expandafter\ifx\csname natexlab\endcsname\relax\def\natexlab#1{#1}\fi
\def\au#1{#1} \def\ed#1{#1} \def\yr#1{#1}\def\at#1{#1}\def\jt#1{\textit{#1}}
  \def\bt#1{#1}\def\bvol#1{\textbf{#1}} \def\vol#1{#1} \def\pg#1{#1}
  \def\publ#1{#1}\def\arxiv#1{#1}\def\org#1{#1}\def\st#1{\textit{#1}}

\bibitem[Ahmed {\em et~al.\/}(2020)Ahmed, Izbassarov, Lu, Tryggvason, Muradoglu
  \& Tammisola]{2020_Ahmed}
{\sc \au{Ahmed, Z.}, \au{Izbassarov, D.}, \au{Lu, J.}, \au{Tryggvason, G.},
  \au{Muradoglu, M.} \& \au{Tammisola, O.}} \yr{2020}  \at{Effects of soluble
  surfactant on lateral migration of a bubble in a pressure driven channel
  flow}.  \jt{Int. J. Multiphase Flow}  \bvol{126},  \pg{103251}.

\bibitem[Barbosa {\em et~al.\/}(2016)Barbosa, Legendre \& Zenit]{2016_Barbosa}
{\sc \au{Barbosa, C.}, \au{Legendre, D.} \& \au{Zenit, R.}} \yr{2016}
  \at{Conditions for the sliding-bouncing transition for the interaction of a
  bubble with an inclined wall}.  \jt{Phys. Rev. Fluids}  \bvol{1},
  \pg{032201}.

\bibitem[Batchelor(1967)]{1967_Batchelor}
{\sc \au{Batchelor, G.~K.}} \yr{1967} {\em An Introduction to Fluid
  Dynamics\/}.  \publ{Cambridge University Press}.

\bibitem[Bonnefis {\em et~al.\/}(2024)Bonnefis, Sierra-Ausin, Fabre \&
  Magnaudet]{2023_Bonnefis_b}
{\sc \au{Bonnefis, P.}, \au{Sierra-Ausin, J.}, \au{Fabre, D.} \& \au{Magnaudet,
  J.}} \yr{2024}  \at{Path instability of deformable bubbles rising in
  {N}ewtonian liquids: {A} linear study}.  \jt{J. Fluid Mech.}  \bvol{980},
  \pg{A19}.

\bibitem[Cai {\em et~al.\/}(2023)Cai, Ju, Chen \& Sun]{2023_Cai}
{\sc \au{Cai, R.}, \au{Ju, E.}, \au{Chen, W.} \& \au{Sun, J.}} \yr{2023}
  \at{Different modes of bubble migration near a vertical wall in pure water}.
  \jt{Korean J. Chem. Eng.}  \bvol{40},  \pg{67--78}.

\bibitem[Cai {\em et~al.\/}(2024)Cai, Sun \& Chen]{2024_Cai}
{\sc \au{Cai, R.}, \au{Sun, J.} \& \au{Chen, W.}} \yr{2024}  \at{Near-wall
  bubble migration and wake structure in viscous liquids}.  \jt{Chem. Eng. Res.
  Des.}  \bvol{202},  \pg{414--428}.

\bibitem[Cano-Lozano {\em et~al.\/}(2016)Cano-Lozano, Mart\'inez-Baz\'an,
  Magnaudet \& Tchoufag]{2016_Cano-Lozano}
{\sc \au{Cano-Lozano, J.~C.}, \au{Mart\'inez-Baz\'an, C.}, \au{Magnaudet, J.}
  \& \au{Tchoufag, J.}} \yr{2016}  \at{Paths and wakes of deformable nearly
  spheroidal rising bubbles close to the transition to path instability}.
  \jt{Phys. Rev. Fluids}  \bvol{1},  \pg{053604}.

\bibitem[Dabiri(2006)]{2006_Dabiri}
{\sc \au{Dabiri, J.~O.}} \yr{2006}  \at{Note on the induced {L}agrangian drift
  and added-mass of a vortex}.  \jt{J. Fluid Mech.}  \bvol{547},
  \pg{105--113}.

\bibitem[Duineveld(1998)]{1998_Duineveld}
{\sc \au{Duineveld, P.~C.}} \yr{1998}  \at{Bouncing and coalescence of bubble
  pairs rising at high {R}eynolds number in pure water or aqueous surfactant
  solutions}.  \jt{Appl. Sci. Res.}  \bvol{58},  \pg{409--439}.

\bibitem[Ellingsen \& Risso(2001)]{2001_Ellingsen}
{\sc \au{Ellingsen, K.} \& \au{Risso, F.}} \yr{2001}  \at{On the rise of an
  ellipsoidal bubble in water: oscillatory paths and liquid-induced velocity}.
  \jt{J. Fluid Mech.}  \bvol{440},  \pg{235--268}.

\bibitem[Estepa-Cantero {\em et~al.\/}(2024)Estepa-Cantero, Mart\'inez-Baz\'an
  \& Bola{\~{n}}os-Jim\'enez]{2024_Estepa}
{\sc \au{Estepa-Cantero, C.}, \au{Mart\'inez-Baz\'an, C.} \&
  \au{Bola{\~{n}}os-Jim\'enez, R.}} \yr{2024}  \at{Bubble rising near a
  vertical wall: Experimental characterization of paths and velocity}.
  \jt{Phys. Fluids}  \bvol{36},  \pg{013304}.

\bibitem[Heydari {\em et~al.\/}(2022)Heydari, Larachi, Taghavi \&
  Bertrand]{2022_Heydari}
{\sc \au{Heydari, N.}, \au{Larachi, F.}, \au{Taghavi, S.~M.} \& \au{Bertrand,
  F.}} \yr{2022}  \at{Three-dimensional analysis of the rising dynamics of
  individual ellipsoidal bubbles in an inclined column}.  \jt{Chem. Eng. Sci.}
  \bvol{258},  \pg{117759}.

\bibitem[van {Hooft} {\em et~al.\/}(2018)van {Hooft}, Popinet, van
  {Heerwaarden}, van~der {Linden}, de~{Roode} \& van~de {Wiel}]{2018_Hooft}
{\sc \au{van {Hooft}, J.~A.}, \au{Popinet, S.}, \au{van {Heerwaarden}, C.~C.},
  \au{van~der {Linden}, S. J.~A.}, \au{de~{Roode}, S.R.} \& \au{van~de {Wiel},
  B. J.~H.}} \yr{2018}  \at{Towards adaptive grids for atmospheric
  boundary-layer simulations}.  \jt{Bound.-Layer Meteorol.}  \bvol{167},
  \pg{421--443}.

\bibitem[Jeong \& Park(2015)]{2015_Jeong}
{\sc \au{Jeong, H.} \& \au{Park, H.}} \yr{2015}  \at{Near-wall rising behaviour
  of a deformable bubble at high {R}eynolds number}.  \jt{J. Fluid Mech.}
  \bvol{771},  \pg{564--594}.

\bibitem[Ju {\em et~al.\/}(2022)Ju, Cai, Sun, Fan, Chen \& Sun]{2022_Ju}
{\sc \au{Ju, E.}, \au{Cai, R.}, \au{Sun, H.}, \au{Fan, Y.}, \au{Chen, W.} \&
  \au{Sun, J.}} \yr{2022}  \at{Dynamic behavior of an ellipsoidal bubble
  contaminated by surfactant near a vertical wall}.  \jt{Korean J. Chem. Eng.}
  \bvol{39},  \pg{1165--1181}.

\bibitem[Khodadadi {\em et~al.\/}(2022)Khodadadi, Samkhaniani, Taleghani,
  Gorji-Bandpy \& Ganji]{2022_Khodadadi}
{\sc \au{Khodadadi, S.}, \au{Samkhaniani, N.}, \au{Taleghani, M.~H.},
  \au{Gorji-Bandpy, M.} \& \au{Ganji, D.~D.}} \yr{2022}  \at{Numerical
  simulation of single bubble motion along inclined walls: {A} comprehensive
  map of outcomes}.  \jt{Ocean Eng.}  \bvol{255},  \pg{111478}.

\bibitem[Klaseboer {\em et~al.\/}(2014)Klaseboer, Manica, Hendrix, Ohl \&
  Chan]{2014_Klaseboer}
{\sc \au{Klaseboer, E.}, \au{Manica, R.}, \au{Hendrix, M. H.~W.}, \au{Ohl,
  C.-D.} \& \au{Chan, D. Y.~C.}} \yr{2014}  \at{A force balance model for the
  motion, impact, and bounce of bubbles}.  \jt{Phys. Fluids}  \bvol{26},
  \pg{092101}.

\bibitem[Kosior {\em et~al.\/}(2014)Kosior, Zawala \& Malysa]{2014_Kosior}
{\sc \au{Kosior, D.}, \au{Zawala, J.} \& \au{Malysa, K.}} \yr{2014}
  \at{Influence of n-octanol on the bubble impact velocity, bouncing and the
  three phase contact formation at hydrophobic solid surfaces}.  \jt{Colloids
  Surf. A Physicochem. Eng. Asp.}  \bvol{441},  \pg{788--795}.

\bibitem[Kusuno \& Sanada(2021)]{2021_Kusuno}
{\sc \au{Kusuno, H.} \& \au{Sanada, T.}} \yr{2021}  \at{Flow structure and
  deformation of two bubbles rising side by side in a quiescent liquid}.
  \jt{Fluids}  \bvol{6},  \pg{390}.

\bibitem[Lamb(1932)]{1932_Lamb}
{\sc \au{Lamb, H.}} \yr{1932} {\em Hydrodynamics\/}.  \publ{Cambridge
  University Press}.

\bibitem[Lee \& Park(2017)]{2017_Lee}
{\sc \au{Lee, J.} \& \au{Park, H.}} \yr{2017}  \at{Wake structures behind an
  oscillating bubble rising close to a vertical wall}.  \jt{Int. J. Multiphase
  Flow}  \bvol{91},  \pg{225--242}.

\bibitem[Lighthill(1956)]{1956_Lighthill}
{\sc \au{Lighthill, M.~J.}} \yr{1956}  \at{Drift}.  \jt{J. Fluid Mech.}
  \bvol{1},  \pg{31--53}.

\bibitem[Lu \& Tryggvason(2013)]{2013_Lu}
{\sc \au{Lu, J.} \& \au{Tryggvason, G.}} \yr{2013}  \at{Dynamics of nearly
  spherical bubbles in a turbulent channel upflow}.  \jt{J. Fluid Mech.}
  \bvol{732},  \pg{166--189}.

\bibitem[Magnaudet \& Mougin(2007)]{2007_Magnaudet}
{\sc \au{Magnaudet, J.} \& \au{Mougin, G.}} \yr{2007}  \at{Wake instability of
  a fixed spheroidal bubble}.  \jt{J. Fluid Mech.}  \bvol{572},  \pg{311--337}.

\bibitem[Michelin {\em et~al.\/}(2019)Michelin, Gallino, Gallaire \&
  Lauga]{2019_Michelin}
{\sc \au{Michelin, S.}, \au{Gallino, G.}, \au{Gallaire, F.} \& \au{Lauga, E.}}
  \yr{2019}  \at{Viscous growth and rebound of a bubble near a rigid surface}.
  \jt{J. Fluid Mech.}  \bvol{860},  \pg{172--199}.

\bibitem[Miloh(1977)]{1977_Miloh}
{\sc \au{Miloh, T.}} \yr{1977}  \at{Hydrodynamics of deformable contiguous
  spherical shapes in an incompressible inviscid fluid}.  \jt{J. Eng. Math.}
  \bvol{11},  \pg{349--372}.

\bibitem[Moore(1965)]{1965_Moore}
{\sc \au{Moore, D.~W.}} \yr{1965}  \at{The velocity of rise of distorted gas
  bubbles in a liquid of small viscosity}.  \jt{J. Fluid Mech.}  \bvol{23},
  \pg{749--766}.

\bibitem[Mougin \& Magnaudet(2006)]{2006_Mougin}
{\sc \au{Mougin, G.} \& \au{Magnaudet, J.}} \yr{2006}  \at{Wake-induced forces
  and torques on a zigzagging/spiralling bubble}.  \jt{J. Fluid Mech.}
  \bvol{567},  \pg{185--194}.

\bibitem[Popinet(2009)]{2009_Popinet}
{\sc \au{Popinet, S.}} \yr{2009}  \at{An accurate adaptive solver for
  surface-tension-driven interfacial flows}.  \jt{J. Comput. Phys.}
  \bvol{228},  \pg{5838--5866}.

\bibitem[Popinet(2015)]{2015_Popinet}
{\sc \au{Popinet, S.}} \yr{2015}  \at{A quadtree-adaptive multigrid solver for
  the {S}erre--{G}reen--{N}aghdi equations}.  \jt{J. Comput. Phys.}
  \bvol{302},  \pg{336--358}.

\bibitem[Sanada {\em et~al.\/}(2009)Sanada, Sato, Shirota \&
  Watanabe]{2009_Sanada}
{\sc \au{Sanada, T.}, \au{Sato, A.}, \au{Shirota, M.} \& \au{Watanabe, M.}}
  \yr{2009}  \at{Motion and coalescence of a pair of bubbles rising side by
  side}.  \jt{Chem. Eng. Sci.}  \bvol{64},  \pg{2659--2671}.

\bibitem[Shi(2024)]{2024_Shi}
{\sc \au{Shi, P.}} \yr{2024}  \at{Reversal of the transverse force on a
  spherical bubble rising close to a vertical wall at moderate-to-high
  {R}eynolds numbers}.  \jt{Phys. Rev. Fluids}  \bvol{9},  \pg{023601}.

\bibitem[Shi {\em et~al.\/}(2020)Shi, Rzehak, Lucas \& Magnaudet]{2020_Shi}
{\sc \au{Shi, P.}, \au{Rzehak, R.}, \au{Lucas, D.} \& \au{Magnaudet, J.}}
  \yr{2020}  \at{Hydrodynamic forces on a clean spherical bubble translating in
  a wall-bounded linear shear flow}.  \jt{Phys. Rev. Fluids}  \bvol{5},
  \pg{073601}.

\bibitem[Sugioka \& Tsukada(2015)]{2015_Sugioka}
{\sc \au{Sugioka, K.~I.} \& \au{Tsukada, T.}} \yr{2015}  \at{Direct numerical
  simulations of drag and lift forces acting on a spherical bubble near a plane
  wall}.  \jt{Int. J. Multiphase Flow}  \bvol{71},  \pg{32--37}.

\bibitem[Takagi \& Matsumoto(2011)]{2011_Takagi}
{\sc \au{Takagi, S.} \& \au{Matsumoto, Y.}} \yr{2011}  \at{Surfactant effects
  on bubble motion and bubbly flows}.  \jt{Annu. Rev. Fluid Mech.}  \bvol{43},
  \pg{615--636}.

\bibitem[Takemura \& Magnaudet(2003)]{2003_Takemura}
{\sc \au{Takemura, F.} \& \au{Magnaudet, J.}} \yr{2003}  \at{The transverse
  force on clean and contaminated bubbles rising near a vertical wall at
  moderate {R}eynolds number}.  \jt{J. Fluid Mech.}  \bvol{495},
  \pg{235--253}.

\bibitem[Takemura {\em et~al.\/}(2002)Takemura, Takagi, Magnaudet \&
  Matsumoto]{2002_Takemura}
{\sc \au{Takemura, F.}, \au{Takagi, S.}, \au{Magnaudet, J.} \& \au{Matsumoto,
  Y.}} \yr{2002}  \at{Drag and lift forces on a bubble rising near a vertical
  wall in a viscous liquid}.  \jt{J. Fluid Mech.}  \bvol{461},  \pg{277--300}.

\bibitem[Tryggvason {\em et~al.\/}(2011)Tryggvason, Scardovelli \&
  Zaleski]{2011_Tryggvason}
{\sc \au{Tryggvason, G.}, \au{Scardovelli, R.} \& \au{Zaleski, S.}} \yr{2011}
  {\em Direct Numerical Simulations of Gas--Liquid Multiphase Flows\/}.
  \publ{Cambridge University Press}.

\bibitem[Tsao \& Koch(1997)]{1997_Tsao}
{\sc \au{Tsao, H.} \& \au{Koch, D.~L.}} \yr{1997}  \at{Observations of high
  {R}eynolds number bubbles interacting with a rigid wall}.  \jt{Phys. Fluids}
  \bvol{9},  \pg{44--56}.

\bibitem[Vasseur \& Cox(1977)]{1977_Vasseur}
{\sc \au{Vasseur, P.} \& \au{Cox, R.~G.}} \yr{1977}  \at{The lateral migration
  of spherical particles sedimenting in a stagnant bounded fluid}.  \jt{J.
  Fluid Mech.}  \bvol{80},  \pg{561--591}.

\bibitem[de~Vries {\em et~al.\/}(2002)de~Vries, Biesheuvel \& van
  Wijngaarden]{2002_devries}
{\sc \au{de~Vries, A. W.~G.}, \au{Biesheuvel, A.} \& \au{van Wijngaarden, L.}}
  \yr{2002}  \at{Notes on the path and wake of a gas bubble rising in pure
  water}.  \jt{Int. J. Multiphase Flow}  \bvol{28},  \pg{1823--1835}.

\bibitem[van {Wijngaarden}(1976)]{1976_Wijngaarden}
{\sc \au{van {Wijngaarden}, L.}} \yr{1976}  \at{Hydrodynamic interaction
  between gas bubbles in liquid}.  \jt{J. Fluid Mech.}  \bvol{77},
  \pg{27--44}.

\bibitem[Yan {\em et~al.\/}(2022)Yan, Zhang, Liao, Zhang, Zhou \&
  Liu]{2022_Yan}
{\sc \au{Yan, H.}, \au{Zhang, H.}, \au{Liao, Y.}, \au{Zhang, H.}, \au{Zhou, P.}
  \& \au{Liu, L.}} \yr{2022}  \at{A single bubble rising in the vicinity of a
  vertical wall: A numerical study based on volume of fluid method}.  \jt{Ocean
  Eng.}  \bvol{263},  \pg{112379}.

\bibitem[Yin \& Koch(2008)]{2008_Yin}
{\sc \au{Yin, X.} \& \au{Koch, D.~L.}} \yr{2008}  \at{Lattice-{B}oltzmann
  simulation of finite {R}eynolds number buoyancy-driven bubbly flows in
  periodic and wall-bounded domains}.  \jt{Phys. Fluids}  \bvol{20},
  \pg{103304}.

\bibitem[Zaruba {\em et~al.\/}(2007)Zaruba, Lucas, Prasser \&
  H{\"o}hne]{2007_Zaruba}
{\sc \au{Zaruba, A.}, \au{Lucas, D.}, \au{Prasser, H.~M.} \& \au{H{\"o}hne,
  T.}} \yr{2007}  \at{Bubble-wall interactions in a vertical gas--liquid flow:
  {B}ouncing, sliding and bubble deformations}.  \jt{Chem. Eng. Sci.}
  \bvol{62},  \pg{1591--1605}.

\bibitem[Zawala \& Dabros(2013)]{2013_Zawala}
{\sc \au{Zawala, J.} \& \au{Dabros, T.}} \yr{2013}  \at{Analysis of energy
  balance during collision of an air bubble with a solid wall}.  \jt{Phys.
  Fluids}  \bvol{25},  \pg{123101}.

\bibitem[Zenit \& Hunt(1999)]{1999_Zenit}
{\sc \au{Zenit, R.} \& \au{Hunt, M.~L.}} \yr{1999}  \at{Mechanics of immersed
  particle collisions}.  \jt{J. Fluids Eng.}  \bvol{121},  \pg{179--184}.

\bibitem[Zenit \& Legendre(2009)]{2009_Zenit}
{\sc \au{Zenit, R.} \& \au{Legendre, D.}} \yr{2009}  \at{The coefficient of
  restitution for air bubbles colliding against solid walls in viscous
  liquids}.  \jt{Phys. Fluids}  \bvol{21},  \pg{083306}.

\bibitem[Zhang {\em et~al.\/}(2021)Zhang, Ni \& Magnaudet]{2021_Zhang}
{\sc \au{Zhang, J.}, \au{Ni, M.} \& \au{Magnaudet, J.}} \yr{2021}
  \at{Three-dimensional dynamics of a pair of deformable bubbles rising
  initially in line. {P}art 1. {M}oderately inertial regimes}.  \jt{J. Fluid
  Mech.}  \bvol{920},  \pg{A16}.

\bibitem[Zhang {\em et~al.\/}(2022)Zhang, Ni \& Magnaudet]{2022_Zhang}
{\sc \au{Zhang, J.}, \au{Ni, M.} \& \au{Magnaudet, J.}} \yr{2022}
  \at{Three-dimensional dynamics of a pair of deformable bubbles rising
  initially in line. {P}art 2. {H}ighly inertial regimes}.  \jt{J. Fluid Mech.}
   \bvol{943},  \pg{A10}.

\bibitem[Zhang {\em et~al.\/}(2020)Zhang, Dabiri, Chen \& You]{2020_Zhang}
{\sc \au{Zhang, Y.}, \au{Dabiri, S.}, \au{Chen, K.} \& \au{You, Y.}} \yr{2020}
  \at{An initially spherical bubble rising near a vertical wall}.  \jt{Int. J.
  Heat Fluid Flow}  \bvol{85},  \pg{108649}.

\end{thebibliography}
\end{document}